\documentclass[12pt]{article}
\pdfoutput=1

\usepackage{cancel,slashed}
\usepackage{bbm}
\usepackage{amssymb}
\usepackage{amsfonts}
\usepackage{amsmath}
\usepackage{graphicx}
\usepackage{subcaption}
\usepackage{cite}

\usepackage{ dsfont }
\usepackage{latexsym}
\usepackage{color}
\usepackage{xypic}
\usepackage{cancel,slashed}
\usepackage[hyperfootnotes=false,linktocpage]{hyperref}
\usepackage{color}
\usepackage{transparent}
\usepackage[hang,flushmargin]{footmisc}

\usepackage{blkarray}
\usepackage{multirow}

\newcommand{\bmat}{\left(\begin{array}}
\newcommand{\emat}{\end{array}\right)}

\def\bZ{\mathbb{Z}}
\def\Z{\mathbb{Z}}
\def\R{\mathbb{R}}
\def\C{\mathbb{C}}

\def\CK {{\cal K}}
\def\a {\alpha}
\def\b {\beta}

\def\ov{\overline}

\def\IM{\text{Im}\,}
\def\RE{\text{Re}\,}
\def\ov{\overline}
\def\1{{\bf 1}}
\def\2{{\bf 2}}
\def\3{{\bf 3}}
\def\4{{\bf 4}}
\def\6{{\bf 6}}

\def\targ#1#2{\genfrac{[}{]}{0pt}{}{#1}{#2}}
\def\targ2#1#2{\genfrac{}{}{0pt}{}{#1}{#2}}

\definecolor{mygr}{rgb}{0,0.6,0}
\definecolor{mygrey}{rgb}{0,0.1,0.2}
\definecolor{myblue}{rgb}{0,0.5,0.9}
\definecolor{myblue2}{rgb}{0,0.5,0.5}
\definecolor{myblue3}{rgb}{0,0.7,0.9}
\definecolor{myblue4}{rgb}{0,0.6,0.6}
\definecolor{myorange}{rgb}{1,0.5,0}
\definecolor{mypurple}{rgb}{0.6,0,1}
\definecolor{mygolden}{rgb}{1,0.8,0.2}
\definecolor{mycyan}{rgb}{0,1,1}
\definecolor{mymagenta}{rgb}{1,0,1}
\definecolor{mykiwi}{rgb}{0.8,1,0.5}
\definecolor{mybrown}{cmyk}{0.14, 0.42, 0.56, 0.2}
\definecolor{myturq}{cmyk}{0.99, 0, 0.2, 0.4}
\definecolor{myaubergine2}{cmyk}{0.4, 0.5, 0, 0.1}
\definecolor{myaubergine}{cmyk}{0.6,0.85,0,0}
\definecolor{CycleGreen}{cmyk}{0.52,0,1,0}
\definecolor{CycleBrown}{cmyk}{0, 0.4, 0.9, 0.2}

\DeclareFontFamily{U}{rcjhbltx}{}
\DeclareFontShape{U}{rcjhbltx}{m}{n}{<->rcjhbltx}{}
\DeclareSymbolFont{hebrewletters}{U}{rcjhbltx}{m}{n}

\DeclareMathSymbol{\lamed}{\mathord}{hebrewletters}{108}
\DeclareMathSymbol{\mem}{\mathord}{hebrewletters}{109}
\DeclareMathSymbol{\ayin}{\mathord}{hebrewletters}{96}
\DeclareMathSymbol{\tsadi}{\mathord}{hebrewletters}{118}
\DeclareMathSymbol{\qof}{\mathord}{hebrewletters}{113}
\DeclareMathSymbol{\resh}{\mathord}{hebrewletters}{114}
\DeclareMathSymbol{\pe}{\mathord}{hebrewletters}{112}
\DeclareMathSymbol{\pesofit}{\mathord}{hebrewletters}{80}
\DeclareMathSymbol{\samekh}{\mathord}{hebrewletters}{115}
\DeclareMathSymbol{\tav}{\mathord}{hebrewletters}{116}
\DeclareMathSymbol{\vav}{\mathord}{hebrewletters}{119}
\DeclareMathSymbol{\het}{\mathord}{hebrewletters}{120}
\DeclareMathSymbol{\yod}{\mathord}{hebrewletters}{121}
\DeclareMathSymbol{\zayin}{\mathord}{hebrewletters}{122}
\DeclareMathSymbol{\alephdot}{\mathord}{hebrewletters}{128}
\DeclareMathSymbol{\tsadisofit}{\mathord}{hebrewletters}{90}
\DeclareMathSymbol{\shin}{\mathord}{hebrewletters}{152}

\def\CN {{\cal N}}
\def\om{{\omega}}
\def\sig{{\sigma}}

\def\be{\begin{equation}}
\def\ee{\end{equation}}
\def\bea{\begin{eqnarray}}
\def\eea{\end{eqnarray}}
\def\bes{\begin{subequations}}
\def\ees{\end{subequations}}
\def\raw{\rightarrow}

\usepackage{multicol}
\usepackage{float}
\usepackage{caption}
\def\mk {{\mathcal K}}

\def\tr {{\tilde{\rho}}}
\def\p {{\partial}}

\def\g {{\gamma}}

\newcommand{\cK}{\mathcal{K}}
\newcommand{\cM}{\mathcal{M}}
\newcommand{\cN}{\mathcal{N}}
\newcommand{\cO}{\mathcal{O}}
\newcommand{\cA}{\mathcal{A}}
\newcommand{\cB}{\mathcal{B}}

\makeatletter
\newsavebox\myboxA
\newsavebox\myboxB
\newlength\mylenA

\newcommand*\xoverline[2][0.75]{%
\sbox{\myboxA}{$\m@th#2$}%
\setbox\myboxB\null
\ht\myboxB=\ht\myboxA%
\dp\myboxB=\dp\myboxA%
\wd\myboxB=#1\wd\myboxA
\sbox\myboxB{$\m@th\overline{\copy\myboxB}$}
\setlength\mylenA{\the\wd\myboxA}
\addtolength\mylenA{-\the\wd\myboxB}%
\ifdim\wd\myboxB<\wd\myboxA%
   \rlap{\hskip 0.5\mylenA\usebox\myboxB}{\usebox\myboxA}%
\else
    \hskip -0.5\mylenA\rlap{\usebox\myboxA}{\hskip 0.5\mylenA\usebox\myboxB}%
\fi}
\makeatother

\begin{document}
\pagestyle{plain}

\makeatletter
\@addtoreset{equation}{section}
\makeatother
\renewcommand{\theequation}{\thesection.\arabic{equation}}

\pagestyle{empty}
\rightline{IFT-UAM/CSIC-20-95}
\vspace{0.5cm}
\begin{center}
\Huge{{Systematics of Type IIA moduli stabilisation}
\\[10mm]}
\normalsize{Fernando Marchesano,$^1$ David Prieto,$^1$ Joan Quirant$^1$ and Pramod Shukla$^2$\\[15mm]}
\small{$^1$ Instituto de F\'{\i}sica Te\'orica UAM-CSIC, Cantoblanco, 28049 Madrid, Spain
\\[1mm]} 
\small{$^2$ ICTP, Strada Costiera 11, Trieste 34151, Italy
\\[8mm]}
\small{\bf Abstract} \\[5mm]
\end{center}
\begin{center}
\begin{minipage}[h]{15.0cm} 

We analyse the flux-induced scalar potential for type IIA orientifolds in the presence of $p$-form, geometric and non-geometric fluxes. Just like in the Calabi--Yau case, the potential presents a bilinear structure, with a factorised dependence on axions and saxions. This feature allows one to perform a systematic search for vacua, which we implement for the case of geometric backgrounds. Guided by stability criteria, we consider configurations with a particular on-shell F-term pattern, and show that no de Sitter extrema are allowed for them. We classify branches of supersymmetric and non-supersymmetric vacua, and argue that the latter are perturbatively stable for a large subset of them. Our solutions reproduce and generalise  previous results in the literature, obtained either from the 4d or 10d viewpoint.

\end{minipage}
\end{center}
\newpage
\setcounter{page}{1}
\pagestyle{plain}
\renewcommand{\thefootnote}{\arabic{footnote}}
\setcounter{footnote}{0}


\tableofcontents

\section{Introduction}
\label{s:intro}

One of the major challenges in the field of string theory is to determine the structure of four-dimensional meta-stable vacua, a.k.a. the string Landscape. Progress in this program has recently taken an interesting turn, as it has been argued that general quantum gravity arguments significantly constrain such structure \cite{Vafa:2005ui,Brennan:2017rbf,Palti:2019pca}. In this context, two of the most dramatic proposals address the difficulties to construct meta-stable de Sitter vacua \cite{Obied:2018sgi,Garg:2018reu,Ooguri:2018wrx} and to achieve separation of scales in anti-de Sitter ones \cite{Lust:2019zwm,Buratti:2020kda}. Remarkably, type IIA flux compactifications have played a key role  in motivating and in testing both Swampland conjectures. To some extent this is because, in appropriate regimes, type IIA moduli stabilisation can be purely addressed at the classical level \cite{Derendinger:2004jn,Villadoro:2005cu,DeWolfe:2005uu,Camara:2005dc}, opening the door for a direct 10d microscopic description of such vacua.  For instance, this feature has recently been exploited in \cite{Font:2019uva,Junghans:2020acz,Marchesano:2020qvg} to test the AdS distance conjectures. 

Despite all these key features, it is fair to say that the general structure of geometric type IIA flux compactifications is less understood than their type IIB counterpart \cite{Grana:2005jc,Douglas:2006es,Becker:2007zj,Denef:2007pq,Ibanez:2012zz}. Part of the problem is all the different kinds of fluxes that are present in the type IIA setup, which, on the other hand, is the peculiarity that permits to stabilise all moduli classically. Traditionally, each kind of flux is treated differently, and as soon as geometric fluxes are introduced the classification of vacua becomes quite involved. 

 The purpose of this paper is to improve this picture by providing a unifying treatment of moduli stabilisation in (massive) type IIA orientifold flux vacua. Our main tool will be the bilinear form of the scalar potential $V = Z^{AB} \rho_A \rho_B$ found in  \cite{Bielleman:2015ina,Carta:2016ynn,Herraez:2018vae}, where $\rho_A$ are axion polynomials with flux-quanta coefficients and the entries of the matrix $Z^{AB}$ only depend on the saxions. While this bilinear structure was originally found for the case of Calabi--Yau compactifications with $p$-form fluxes, building on \cite{Gao:2017gxk} we show that it can be extended to include the presence of geometric and non-geometric fluxes, even when these fluxes generate both an F-term and a D-term potential. 
 
 With this form of the flux potential, one may perform a systematic search for vacua, as already carried out for the Calabi--Yau case \cite{Escobar:2018tiu,Escobar:2018rna,Marchesano:2019hfb}. We do so for the case of orientifold compactifications with $p$-form and geometric fluxes, which are one of the main sources of classical AdS$_4$ and dS$_4$ backgrounds in string theory, and have already provided crucial information regarding swampland criteria. On the one hand, the microscopic 10d description of AdS$_4$ geometric flux vacua has been discussed in several instances  \cite{House:2005yc,Grana:2006kf,Aldazabal:2007sn,Koerber:2008rx,Caviezel:2008ik,Koerber:2010rn}. On the other hand, they have provided several no-go results on de Sitter solutions \cite{Hertzberg:2007wc,Haque:2008jz,Caviezel:2008tf,Flauger:2008ad,Danielsson:2009ff,Danielsson:2010bc,Danielsson:2011au}, as well as examples of unstable de Sitter extrema that have served to refine the original de Sitter conjecture \cite{Andriot:2018wzk}.  Therefore, it is expected that a global, more exhaustive description of this class of vacua  and a systematic understanding of their properties  leads to further tests, and perhaps even refinements, of the de Sitter and AdS distance conjectures.
  
To perform our search for vacua we consider a certain pattern of on-shell F-terms, that is then translated into an Ansatz. Even if this F-term pattern is motivated from general stability criteria for de Sitter vacua \cite{GomezReino:2006dk,GomezReino:2006wv,GomezReino:2007qi,Covi:2008ea,Covi:2008zu}, one can show that de Sitter extrema are incompatible with such F-terms, obtaining a new kind of no-go result. Compactifications to AdS$_4$ are on the other hand allowed, and using our Ansatz we find both a supersymmetric  and a non-supersymmetric branch of vacua, intersecting at one point. In some cases we can check explicitly the perturbative stability of the non-SUSY AdS$_4$ branch, finding that the vacua are stable for a large region of the parameter space of our Ansatz, and even free of tachyons for a large subregion. We finally comment on the 10d description of this set of vacua.

The paper is organised as follows. In section \ref{s:IIAorientifold} we consider the classical F-term and D-term potential of type IIA  compactifications with all kind of fluxes and express both potentials in a bilinear form. In section \ref{s:fluxpot} we propose an F-term pattern to avoid tachyons in de Sitter vacua, and build a general Ansatz from it. We also describe the flux invariants present in this class of compactifications. In section \ref{s:geovacua} we apply our results to configurations with $p$-form and geometric fluxes, in order to classify their different extrema. We find two different branches, that contain several previous results in the literature. In section \ref{s:stabalidity} we discuss which of these extrema are perturbatively stable, as well as their 10d description. We draw our conclusions in section \ref{s:conclu}. 

Some technical details have been relegated to the Appendices. Appendix \ref{ap:conv} contains several aspects regarding NS fluxes and flux-axion polynomials. Appendix \ref{ap:curvature} develops the computations motivating our F-term Ansatz. Appendix \ref{ap:Hessian} contains the computation of the Hessian for geometric flux extrema.


\section{The Type IIA general flux potential}
\label{s:IIAorientifold}

Type IIA compactifications with orientifolds and fluxes represent a particularly interesting corner of the string landscape, as already the classical potential generated by $p$-form fluxes suffices to stabilise all moduli \cite{DeWolfe:2005uu,Camara:2005dc}. Even so, as pointed out in \cite{Shelton:2005cf} one may consider a larger set of NS fluxes for this class of compactifications, related to each other by T-duality. Taking them into account results into a richer scalar potential, as analysed in \cite{Aldazabal:2006up,Shelton:2006fd,Micu:2007rd,Ihl:2007ah,Wecht:2007wu,Robbins:2007yv}. In this section we consider the scalar potential obtained from the set of geometric and non-geometric fluxes defined over a Calabi--Yau manifold. As shown in \cite{Gao:2017gxk}, such a potential can be expressed in a quite compact form, reminiscent of the bilinear expression introduced in \cite{Bielleman:2015ina,Herraez:2018vae} for the case with $p$-form fluxes. This last form is particularly useful to study the vacua of the scalar potential, as demonstrated in \cite{Escobar:2018tiu,Escobar:2018rna,Marchesano:2019hfb} (see also \cite{Valenzuela:2016yny}). Therefore in the following we will adapt the results of \cite{Gao:2017gxk} to rewrite the potential in a bilinear form, in order to perform our moduli stabilisation analysis in subsequent sections.

\subsection{Type IIA orientifolds with general fluxes}

Let us consider type IIA string theory compactified on an orientifold of $X_4 \times X_6$ with $X_6$ a compact Calabi--Yau three-fold. We take the standard orientifold quotient by $\Omega_p (-)^{F_L} {\cal R}$ \cite{Blumenhagen:2005mu,Blumenhagen:2006ci,Marchesano:2007de,Ibanez:2012zz},\footnote{Here $\Omega_p$ the worldsheet parity reversal operator and  ${F_L}$ spacetime fermion number for the left movers.}  with ${\mathcal R}$ an involution of the Calabi--Yau metric acting on the K\"ahler 2-form $J$ and the holomorphic 3-form $\Omega$ as ${\cal R}(J) = - J$ and ${\cal R} (\Omega) = e^{2i\theta} \ov \Omega$, respectively.

In the absence of background fluxes and neglecting worldsheet and D-brane instanton effects, dimensional reduction to 4d of the closed string sector  yields several massless chiral fields, whose scalar components are described in terms of the Calabi--Yau harmonic forms and their ${\cal R}$ eigenvalue  \cite{Grimm:2004ua}. We summarise in table \ref{tab_1} the various  ${\cal R}$-even and odd cohomology groups of $X_6$, together with their harmonic representatives.
\begin{table}[H]
\begin{center}
\begin{tabular}{|c|| c| c| c| c| c| c|} 
\hline
Cohomology group & $H^{1,1}_+$ & $H^{1,1}_-$ & $H^{2,2}_+$ & $H^{2,2}_-$ & $H^{3}_+$ & $H^{3}_-$ \\
\hline\hline
 Dimension & $h^{1,1}_+$ & $h^{1,1}_-$ & $h^{1,1}_-$ & $h^{1,1}_+$ & $h^{2,1}+ 1$ & $h^{2,1} + 1$ \\
 Basis & $\varpi_\alpha$ & $\omega_a$ & $\tilde\omega^a$ & $\tilde\varpi^\alpha$ & $\alpha_I$ , $\beta^\Lambda$ & $\beta^J$ , $\alpha_\Sigma$ \\
 \hline
\end{tabular}
\end{center}
\caption{Representation of various harmonic forms and their counting.}
\label{tab_1}
\end{table}
\noindent
Moreover, we consider the basis of harmonic representatives to be quantised in units of the string length $\ell_s = 2 \pi \sqrt{\alpha'}$, such that they satisfy
\begin{align}
\label{eq:intersectionBases}
&  \frac{1}{\ell_s^{6}} \int_{X_6} \Phi_6 = 1\, , \quad  \frac{1}{\ell_s^{6}} \int_{X_6} \om_a \wedge \om_b \wedge \om_c= {\cal K}_{abc} \, , \quad \frac{1}{\ell_s^{6}} \int_{X_6} \om_a \wedge \varpi_\alpha \wedge \varpi_\beta = \hat{\cal K}_{a \alpha \beta}\, ,\\ \nonumber
& \frac{1}{\ell_s^{6}}  \int_{X_6} \om_a \wedge \tilde{\om}^b= \delta_a{}^b , \quad   \frac{1}{\ell_s^{6}}  \int_{X_6} \varpi_\alpha \wedge \tilde{\varpi}^\beta = {\delta}_\alpha{}^\beta , \quad  \frac{1}{\ell_s^{6}} \int_{X_6} \alpha_I \wedge \beta^J=\delta_I{}^J, \quad  \frac{1}{\ell_s^{6}}  \int_{X_6} \alpha_\Sigma \wedge \beta^\Lambda =\delta_\Sigma{}^\Lambda ,
\end{align}
where ${\cal K}_{abc}, \hat{\cal K}_{a \alpha \beta}$ are triple intersection numbers and we have introduced the normalised volume form $\Phi_6$. We define the complexified K\"ahler moduli $T^a = b^a + it^a$  through 
\begin{equation}
J_c \equiv B + i\, e^{\frac{\phi}{2}} J = \left( b^a + i t^a\right) \om_a \, , 
\end{equation} 
where $J$ is expressed in the Einstein frame and $\phi$ is the 10d dilaton. The kinetic terms for these moduli is encoded in their K\"ahler potential
\be
K_K \,  = \, -{\rm log} \left(\frac{i}{6} \CK_{abc} (T^a - \bar{T}^a)(T^b - \bar{T}^b)(T^c - \bar{T}^c) \right) \, = \,  -{\rm log} \left(\frac{4}{3} \cK\right) \, ,
\label{KK}
\ee
where $\cK = \cK_{abc} t^at^bt^c = 6 {\rm Vol}_{X_6} = \frac{3}{4} {\cal G}_T$ is homogeneous of degree three on the $t^a$.

The remaining moduli of the compactification are a combination of complex structure moduli and axions arising from the RR three-form potential $C_3$. To define them one first expands the CY three-form as
\be
\Omega = {\cal Z}^\kappa \alpha_\kappa - {\cal F}_\lambda \beta^\lambda\, ,
\ee
where $(\alpha_\kappa, \beta^\lambda) \in H_3({\cal M}_6, \Z)$ is a symplectic basis of three-forms. The orientifold projection  decomposes this basis into ${\cal R}$-even $(\alpha_K, \beta^\Lambda) \in H_+^3$ and ${\cal R}$-odd 3-forms $(\beta^K, \alpha_\Lambda) \in H_-^3$, and eliminates half of the degrees of freedom of the original complex periods of $\Omega$. 
Then one defines the complexified 3-form $\Omega_c$ as
\begin{equation}
\Omega_c \equiv C_3 + i \, \RE ({\cal C} \Omega)\, ,
\end{equation}
where ${\cal C} \equiv  e^{- \phi- i \theta} e^{\frac{1}{2}(K_{cs} - K_T)}$ and $K_{cs} = - \log \left( \frac{1}{i\ell_s^{6}} \int_{X_6} \Omega \wedge  \ov \Omega \right)$. Finally, the moduli including the complex structure are defined as:
\begin{equation}\label{cpxmoduli}
N^K = \xi^K + i n^K = \ell_s^{-3} \int_{X_6} \Omega_c \wedge \beta^K, \qquad  U_{\Lambda} = \xi_\Lambda + i u_\Lambda =  \ell_s^{-3} \int_{X_6} \Omega_c \wedge \alpha_\Lambda.
\end{equation}
Their kinetic terms are given in terms of the following piece of the K\"ahler potential:
\begin{equation}\label{KQ}
 K_Q = -2 \log \left( \frac{1}{4} \RE({\cal C}{\cal Z}^K) \IM({\cal C} {\cal F}_K) - \frac{1}{4} \IM({\cal C} {\cal Z}^\Lambda) \RE({\cal C} {\cal F}_\Lambda) \right) = 4D\, ,
\end{equation}
where $D$ is the four-dimensional dilaton $e^{D} \equiv \frac{e^{\phi}}{{\rm Vol}_{X_6}^{1/2}}$. The periods ${\cal F}_K$ and ${\cal F}_\Lambda$ are homogeneous functions of degree one in ${\cal Z}^K$ and ${\cal Z}^\Lambda$, and so the function ${\cal G}_Q= e^{-K_Q/2}$ is homogeneous of degree two in $n^K$, $u_{\Lambda}$. These moduli are redefined in the presence of D6-brane moduli, and so is the K\"ahler potential \eqref{KQ} \cite{Grimm:2011dx,Kerstan:2011dy,Carta:2016ynn,Herraez:2018vae}. For simplicity, we will not consider compactifications with D6-brane moduli in the following.

In addition to the spectrum of chiral multiplets, vector multiplets arise from the dimensional reduction of the closed string sector. More precisely, dimensionally reducing the RR potentials yields
\be
C_3 = \xi^K  \alpha_K +  \xi_\Lambda  \beta^ \Lambda + A^\alpha \varpi_\alpha, \qquad C_5 =  C_{2\, J} \beta^J + C_2^\Sigma \a_\Sigma  + A_\alpha \tilde\varpi^\alpha \, ,
\label{C3decomp}
\ee
where $C_{2\, J}$, $C_2^\Sigma$ are the 4d two-forms dual to the axions $\xi^K$, $\xi_\Lambda$, respectively. The vectors $A^\alpha$ represent each of the $U(1)$ gauge generators of the closed string sector, with gauge kinetic function
\be
\label{eq:fg}
2 f_{\alpha\beta} = i\, \hat{\cal K}_{a\alpha\beta} \, T^a\, ,
\ee
 and their magnetic duals correspond to \cite{Kerstan:2011dy}
\be
d \left(A_\a - \hat{\cal K}_{a\alpha\beta} b^a A^\beta\right) = - 2 \RE f_{\alpha\beta} *_4 dA^\beta - 2\IM f_{\a\b} dA^\beta \, .
\ee

\subsubsection*{The flux superpotential}

The flux superpotential including RR and geometric and non-geometric NS fluxes is described in terms of a twisted differential operator \cite{Shelton:2006fd}
\be
\label{eq:twistedD}
{\cal D} = d + H \wedge  +\  f \triangleleft  +\ Q \triangleright  +\  R\, \bullet \, ,
\ee
where $H$ is the NS three-form flux, $f$ encodes the geometric fluxes, $Q$ that of globally-non-geometric fluxes and $R$ is the locally-non-geometric fluxes, see e.g. \cite{Wecht:2007wu,Plauschinn:2018wbo} for more details. The action of various fluxes appearing in ${\cal D}$ is such that for an arbitrary $p$-form $A_p$, the pieces $H\wedge A_p$, $f \triangleleft A_p$, $Q \triangleright A_p$ and $R \bullet A_p$ denote a $(p+3)$, $(p+1)$, $(p-1)$ and $(p-3)$-form respectively. We describe their action on the basis of harmonic forms in Appendix \ref{ap:conv}. In addition, the internal RR fluxes can be gathered in a single polyform
\be
F_{RR} = F_0 + F_2 + F_4 + F_6 \,.
\ee
Given these definitions, the flux-generated superpotential  reads $W = W_{\rm RR} + W_{\rm NS}$ \cite{Shelton:2006fd,Aldazabal:2006up}
\be
\label{eq:Wrrnsns}
  W_{\rm RR}=  \frac{1}{\ell_s^6} \int_{X_6} e^{-J_c} \wedge F_{RR}\, , \qquad  W_{\rm NS} =   \frac{1}{\ell_s^6}  \int_{X_6} \Omega_c \wedge {\cal D}\left( e^{J_c} \right)\,.
\ee
Expanding the $p$-form field strengths in the basis of quantised forms
\bea
\label{eq:flux-components}
& & \hskip-1cm F_0 =  - m, \quad F_2 = m^a\, \om_a, \quad F_4 = - e_a\, \tilde{\om}^a, \quad F_6 = e_0\, \Phi_6 \, , \quad   H = h_K \beta^K - h^\Lambda \a_\Lambda\, ,
\eea
and using the action of the NS fluxes on such a basis as given in \eqref{eq:fluxActions0}, one obtains the following expressions \cite{Shelton:2006fd,Aldazabal:2006up,Micu:2007rd,Wecht:2007wu,Ihl:2007ah}
\bea
\label{eq:Wgen}
\ell_s W_{\rm RR} &= &e_0 +  e_aT^a + \frac{1}{2}\, {\cal K}_{abc}  m^a T^b T^c  + \frac{m}{6}\, {\cal K}_{abc}\, T^a T^bT^c \, , \\
\ell_s W_{\rm NS} &= & U^\mu \Bigl[ h_\mu + f_{a\mu} T^a + \frac{1}{2} {\cal K}_{abc} \, T^b \, T^c \, Q^a{}_\mu + \frac{1}{6}\, {\cal K}_{abc} T^a T^b T^c \, R_\mu \Bigr] \, ,
\label{eq:WgenNS}
\eea
where for simplicity we have collected both sets of moduli $(N^K, U_\Lambda)$ into $U^\mu$, and modified the definition of the fluxes accordingly. Here $e_0, e_a, m^a, m, h_\mu,  f_{a\mu}, Q^a{}_\mu,  R_\mu$ are all integers.

\subsection{The F-term flux potential}

Under the assumption that background fluxes do not affect the K\"ahler potential pieces \eqref{KK} and \eqref{KQ},\footnote{The validity of this assumption should not be taken for granted and will depend on the particular class of vacua. The results in \cite{Junghans:2020acz,Buratti:2020kda,Marchesano:2020qvg} suggest that it is valid in the presence of only $p$-form fluxes $F_{\rm RR}$, $H$. However,  \cite{Font:2019uva} gives an example of compactification with metric fluxes in which the naive KK scale is heavily corrected by fluxes, and so should be the K\"ahler potential.} one can easily compute the F-term flux potential for closed string moduli via the standard supergravity expression
\be
\label{eq:VFgen}
\kappa_4^2\, V_F =  e^K \left(K^{{\cal A}\ov{\cal B}}\, D_{\cal A} W \, \ov{D}_{\ov{\cal B}^\prime} \ov{W} - 3 \, |W|^2\right),
\ee
where the index ${\cal A} = \{a, \mu\}$ runs over all  moduli. As in \cite{Bielleman:2015ina,Herraez:2018vae}, one can show that this F-term potential displays a bilinear structure of the form
\begin{equation}\label{VF}
\kappa_4^2\, V_F  = {\rho}_\cA \, Z^{\cA\cB} \, {\rho}_\cB\, , 
\end{equation}  
where the matrix entries $Z^{\cA\cB}$ only depend on the saxions $\{t^a, n^\mu \}$, while the ${\rho}_\cA$ only depend on the flux quanta and the axions $\{b^a, \xi^\mu\}$. Indeed, one can easily rewrite the results in \cite{Gao:2017gxk} to fit the above expression, obtaining the following result.

The set of axion polynomials with flux-quanta coefficients are
\begin{equation}
    \rho_\cA=\{\rho_0,\rho_a,\tilde{\rho}^a,\tilde{\rho},\rho_\mu,\rho_{a\mu},\tilde{\rho}^a_\mu ,\tilde{\rho}_\mu \}\, ,
    \label{rhos}
\end{equation}
and are defined as
\bes
\label{RRrhos}
\begin{align}
  \ell_s  \rho_0&=e_0+e_ab^a+\frac{1}{2}\mathcal{K}_{abc}m^ab^bb^c+\frac{m}{6}\mathcal{K}_{abc}b^ab^bb^c+\rho_\mu\xi^\mu\, , \label{eq: rho0}\\
 \ell_s   \rho_a&=e_a+\mathcal{K}_{abc}m^bb^c+\frac{m}{2}\mathcal{K}_{abc}b^bb^c+\rho_{a\mu}\xi^\mu \, ,  \label{eq: rho_a}\\
  \ell_s  \tilde{\rho}^a&=m^a+m b^a + \tilde{\rho}^a_\mu\xi^\mu \, ,  \label{eq: rho^a}\\
 \ell_s   \tilde{\rho}&=m+\tilde{\rho}_\mu\xi^\mu \, ,   \label{eq: rhom}
\end{align}   
\ees 
and
\bes
\label{NSrhos}
\begin{align}    
\ell_s    \rho_\mu&=h_\mu+f_{a\mu}b^a+\frac{1}{2}\mathcal{K}_{abc}b^bb^cQ_\mu^a+\frac{1}{6}\mathcal{K}_{abc}b^ab^bb^cR_\mu \, , \\
 \ell_s   \rho_{a\mu}&=f_{a\mu}+\mathcal{K}_{abc}b^bQ^c_\mu+\frac{1}{2}\mathcal{K}_{abc}b^bb^cR_\mu \, ,  \label{eq: rho_ak} \\
 \ell_s   \tilde{\rho}^a_\mu &=Q^a_\mu+b^aR_\mu \, , \\
 \ell_s   \tilde{\rho}_\mu &=R_\mu \, .
\end{align}
\ees
The polynomials \eqref{NSrhos} are mostly new with respect to the Calabi--Yau case with $p$-form fluxes, as they highly depend on the presence of geometric and non-geometric fluxes. As in \cite{Herraez:2018vae}, both \eqref{RRrhos} and \eqref{NSrhos} have the interpretation of invariants under the discrete shift symmetries of the combined superpotential $W = W_{\rm RR} + W_{\rm NS}$. This invariance is more manifest by writing $\ell_s \rho_\cA  = {\cal R}_\cA{}^\cB q_\cB$, where $q_\cA = \left\{e_0, \, e_b, \,  m^b, \, m, \, h_\mu, \, f_{b\mu}, \, Q^b{}_\mu, \, R_\mu \right\}$ encodes the flux quanta of the compactification and
\be
\label{eq:invRmat}
{\cal R} = \begin{bmatrix}
    {\cal R}_0  \quad & \qquad {\cal R}_0 \, \, \, \xi^\mu \,  \\
  0  \quad  & \qquad {\cal R}_0 \,\, \, \delta_\nu^\mu
\end{bmatrix}\, , \quad 
 {\cal R}_0 = \begin{bmatrix}
 1 & \quad b^b & \quad \frac{1}{2} \, {\cal K}_{abc} \, b^a \, b^c & \quad \frac{1}{6}\, {\cal K}_{abc} \, b^a \, b^b \, b^c \\
 0 & \quad \delta_a^b & \quad {\cal K}_{abc} \, b^c & \quad \frac{1}{2} \, {\cal K}_{abc} \, b^b \, b^c \\
 0 & \quad 0 & \quad \delta_b^a & \quad b^a\\
 0 & \quad 0 & 0 & \quad 1 \\
\end{bmatrix}\, ,
\ee
is an axion-dependent upper triangular  matrix, see Appendix \ref{ap:conv} for details. Including curvature corrections will modify ${\cal R}_0$, such that discrete shift symmetries become manifest, and shifting an axion by a unit period can be compensated by an integer shift of $q_\cA$ \cite{Escobar:2018rna}.

As for the bilinear form $Z$, one finds the following expression
\be
\label{eq:Z-matrix}
Z^{{\cal A}{\cal B}} =  e^K \, \begin{bmatrix}
   {\bf G}  \quad & \, \, {\cal O} \\
   {\cal O}^{\, t} \quad  & \, \, {\bf C}
\end{bmatrix}\, ,
\ee
where
\begin{equation}
    {\bf G} =\left(\begin{array}{cccc}
         4 & 0 & 0 & 0 \\
         0 & g^{ab} & 0 & 0\\
         0 & 0 & \frac{4\mathcal{K}^2}{9}g_{ab} & 0 \\
         0 & 0 & 0 & \frac{\mathcal{K}^2}{9}
    \end{array}
    \right)\, , \quad 
        \mathcal{O} =\left(\begin{array}{cccc}
         0 & 0 & 0 & -\frac{2\mathcal{K}}{3}u^\nu \\
         0 & 0 &  \frac{2\mathcal{K}}{3}u^\nu \delta^a_b & 0\\
         0 & -\frac{2\mathcal{K}}{3}u^\nu\delta^b_a & 0 & 0 \\
         \frac{2\mathcal{K}}{3}u^\nu & 0 & 0 & 0
    \end{array}
    \right)\, ,
    \label{eq:GOmatrix}
\end{equation}
\begin{equation}
    {\bf C} =\left(\begin{array}{cccc}
         c^{\mu\nu} & 0 & -\tilde{c}^{\mu\nu}\frac{\mathcal{K}_b}{2} & 0 \\
         0 & \tilde{c}^{\mu\nu}t^at^b+ g^{ab}u^\mu u^\nu  &  0 & -\tilde{c}^{\mu\nu}t^a\frac{\mathcal{K}}{6}\\
         -\tilde{c}^{\mu\nu}\frac{\mathcal{K}_a}{2} & 0 & \frac{1}{4}\tilde{c}^{\mu\nu}\mathcal{K}_a\mathcal{K}_b+\frac{4\mathcal{K}^2}{9}g_{ab}u^\mu u^\nu  & 0 \\
         0 & -\tilde{c}^{\mu\nu}t^b\frac{\mathcal{K}}{6} & 0 & \frac{\mathcal{K}^2}{36}c^{\mu\nu}
    \end{array}
    \right)\, .
\end{equation}
Here $K = K_K + K_Q$, $g_{ab} = \frac{1}{4} \partial_{t^a} \partial_{t^b} K_K\equiv \frac{1}{4}\p_a\p_b K_K$, and $c_{\mu\nu} = \frac{1}{4} \partial_{u^\mu}\partial_{u^\nu} K_Q\equiv \frac{1}{4}\p_\mu\p_\nu K_Q$, while upper indices denote their inverses. Also $u^\mu = \IM U^\mu =(n^K, u_\Lambda)$ stands for the complex structure saxions, and we have defined  $\mk_{a}=\mk_{abc}t^bt^c$ and $\tilde{c}^{\mu\nu}=c^{\mu\nu}-4u^\mu u^\nu $. 

Compared to the Calabi--Yau case of \cite{Herraez:2018vae,Marchesano:2019hfb} the matrices {\bf C} and ${\cal O}$ are more involved, again due to the presence of geometric and non-geometric fluxes. Interestingly, the off-diagonal matrix ${\cal O}$ has the same source as in the Calabi--Yau case, namely the contribution from the tension of the localised sources after taking into account  tadpole cancellation. Indeed, the contribution of background fluxes to the D6-brane tadpole is given by \cite{Aldazabal:2006up}
\be
{\cal D} F_{RR} = - \left(m h_\mu  -  m^a f_{a \mu} +  e_aQ^a{}_\mu  -  e_0 R_\mu  \right)\, \beta^\mu \, ,
\label{DFtadpole}
\ee
which can be easily expressed in terms of the $\rho_\cA$.
The corresponding absence of D6-branes needed to cancel such tadpole then translates into the following piece of the potential 
\be
\kappa_4^2 V_{\rm loc} = \frac{4}{3} e^K {\cal K}\, u^\mu  \left( \tilde\rho  \rho_\mu -  \tilde\rho^a  \rho_{a\mu}  +  \rho_a \tilde\rho^a{}_\mu - \rho_0 \, \tilde\rho_\mu \right)\, ,
\ee 
which is nothing but the said off-diagonal contribution.

Putting all this together, the final expression for the F-term potential reads
\begin{align}
   \kappa_4^2 V_F =\, &e^K\left[4\rho_0^2+g^{ab}\rho_a\rho_b+\frac{4\mathcal{K}^2}{9}g_{ab}\tilde{\rho}^a\tilde{\rho}^b+\frac{\mathcal{K}^2}{9}\tilde{\rho}^2+c^{\mu\nu}\rho_\mu\rho_\nu+\left(\tilde{c}^{\mu\nu}t^at^b+g^{ab}u^\mu u^\nu \right)\rho_{a\mu}\rho_{b\nu}\right.\nonumber\\
    &+\left(\tilde{c}^{\mu\nu}\frac{\mathcal{K}_a}{2}\frac{\mathcal{K}_b}{2}+\frac{4\mathcal{K}^2}{9}g_ {ab}u^\mu u^\nu \right)\tilde{\rho}^a_\mu\tilde{\rho}^b_\nu+\frac{\mathcal{K}^2}{36}c^{\mu\nu}\tilde{\rho}_\mu\tilde{\rho}_\nu-\frac{4\mathcal{K}}{3}u^\nu \rho_0\tilde{\rho}_\nu+\frac{4\mathcal{K}}{3}u^\nu \rho_a\tilde{\rho}^a_\nu\nonumber\\
    &\left.-\frac{4\mathcal{K}}{3}u^\nu \tilde{\rho}^a\rho_{a\nu}+\frac{4\mathcal{K}}{3}u^\nu \tilde{\rho}\rho_\nu-\tilde{c}^{\mu\nu}\mathcal{K}_a\rho_\mu\tilde{\rho}^a_\nu-\tilde{c}^{\mu\nu}t^a\frac{\mathcal{K}}{3}\rho_{a\mu}\tilde{\rho}_\nu\right]\, .
    \label{eq:potential}
\end{align}
This expression generalises the result of \cite{Herraez:2018vae} and can be easily connected to other known formulations of (non-)geometric potentials in the type IIA literature, e.g. \cite{Villadoro:2005cu,Flauger:2008ad,Blumenhagen:2013hva,Shukla:2019akv}.

\subsection{The D-term flux potential}

In the presence of a non-trivial even cohomology group $H_+^{1,1}$, $U(1)$ gauge symmetries arise from the closed string sector of the compactification. In addition, as pointed out in \cite{Ihl:2007ah,Robbins:2007yv}, the presence of geometric and non-geometric fluxes will generate a D-term contribution to the scalar potential. This can be computed as
\be
\label{eq:VDgen}
V_D = \frac{1}{2}  \left( {\rm Re} f\right)^{-1\: \alpha\beta}\, D_\alpha \, D_\beta\, ,
\ee
where $D_\alpha$ is the $D$-term for the $U(1)$ gauge group corresponding to a 1-form potential $A^\alpha$ 
\be
D_\alpha = i \partial_{{\cal A}} K \, \delta_\alpha \varphi^{\cal A} + \zeta_\alpha \,,
\ee
where $\delta_\alpha \varphi^{\cal A}$ is the variation of the scalar field $\varphi^{\cal A}$ under a gauge transformation, and $\zeta_\alpha$ is the corresponding Fayet-Iliopoulos term.
In order to find the explicit expression of the D-term potential we perform a gauge transformation on the gauge bosons in \eqref{C3decomp} 
\be
A^\alpha\, , A_\alpha \quad \longrightarrow  \quad A^\alpha + d \lambda^\alpha\, ,\, A_\alpha  \to A_\alpha + d \lambda_\alpha\, .
\ee
The transformation of the RR $p$-form potential $C_{RR}  \equiv  C_1 + C_3 + C_5 + \dots $ can then be given in terms of the twisted differential ${\cal D}$ given in  \eqref{eq:twistedD}
\bea
\label{eq:C3change}
C_{RR}   & \longrightarrow &  C_{RR} + {\cal D}\left(\lambda^\alpha \, \varpi_\alpha + \lambda_\alpha \, \tilde\varpi^{\alpha} \right) \\
& = & \left(\xi^K + \lambda^\alpha \, \hat f_{\alpha}{}^K + \lambda_\alpha\, \hat{Q}^{\alpha K}\right) \, \alpha_K -  \left(\xi_\Lambda  + \lambda^\alpha \, \hat f_{\alpha\Lambda} + \lambda_\alpha\, \hat{Q}^{\alpha}{}_\Lambda\right)  \beta^ \Lambda + \dots \nonumber
\eea
where we have used the flux actions given in (\ref{eq:fluxActions0}), with $\hat f_{\alpha}{}^K$, $\hat{Q}^{\alpha K}$, $\hat f_{\alpha\Lambda}$, $\hat{Q}^{\alpha}{}_\Lambda$ integers. This transformation shows that the scalar fields $\xi^K$, $\xi_\Lambda$ are not invariant under the gauge transformation, leading to the following shift in the ${\cal N} = 1$ coordinates $U^\mu = (N^K, U_\Lambda)$,
\be 
\delta U^\mu =  \lambda^\alpha\, \hat f_{\alpha}{}^\mu +  \lambda_\alpha \, \hat{Q}^{\alpha \mu}\, ,
\label{transU}
\ee
where we have again unified the NS fluxes under the index $\mu$. Note that due to the Bianchi identities \eqref{eq:bianchids2} only the combinations of fields  $U^\mu$ invariant under \eqref{transU} appear in the superpotential and, as a result, the Fayet-Iliopoulos terms vanish.  Interpreting \eqref{transU} as gaugings of the U(1) gauge fields and their magnetic duals one obtains the D-terms
\be
D_\alpha = \frac{1}{2} \partial_\mu K \, \left(\hat f_\alpha{}^\mu +\hat{\cal K}_{a\alpha\beta} b^a \hat{Q}^{\beta \mu}  \right) \, , \qquad   D^\alpha =\frac{1}{2}  \partial_\mu K \, \hat{Q}^{\alpha \mu} \, .
\ee
Taking into account the kinetic couplings \eqref{eq:fg} we end up with the following D-term scalar potential 
\be
\label{eq:DtermGen}
V_D =-\frac{1}{4}  \partial_\mu K \partial_\nu K \biggl({\rm Im}\, \hat{\cal K}^{-1\: \alpha\beta} \left(\hat f_\alpha{}^\mu +\hat{\cal K}_{a\alpha\gamma} b^a \hat{Q}^{\gamma \mu}  \right) \left(\hat f_\beta{}^\nu +\hat{\cal K}_{c\beta\delta} b^c \hat{Q}^{\delta \nu}  \right)
 +  {\rm Im}\, \hat{\cal K}_{\alpha\beta}  \hat{Q}^{\alpha \mu} \, \hat{Q}^{\beta \nu} \biggr) \, ,
\ee
where $\hat{\cal K}_{\alpha\beta} = \hat{\cal K}_{a\alpha\beta} \, T^a$. Alternatively, one may obtain the same potential by following the tensor multiplet analysis of \cite{Grimm:2004uq,Louis:2004xi}.\footnote{This result is different from the type IIA D-term potential of \cite{Robbins:2007yv}, and recovers the expected discrete gauge symmetries related to $b$-field shifts. The same strategy can be applied to type IIB setups with non-geometric fluxes, recovering the full scalar  obtained by DFT dimensional reduction in   \cite{Blumenhagen:2015lta}.}

Finally, one can rewrite this expression in a bilinear form similar to \eqref{VF} by defining the following flux-axion polynomials
\be
\label{eq:NSorbitsNew2}
\ell_s \hat{\rho}_\alpha{}^\mu = \hat f_\alpha{}^\mu + \hat{\cal K}_{a\alpha\beta}\, b^a \,  \hat{Q}^{\beta \mu}\,,\qquad
 \ell_s\tilde\rho^{\alpha \mu} = \hat{Q}^{\alpha \mu} \, ,
\ee
so that one has 
\bea\nonumber
 \kappa_4^2 V_D& = &\frac{1}{4}
\begin{bmatrix}
\hat{\rho}_\alpha{}^\mu & \, \,  \tilde\rho^{\alpha \mu} \\
\end{bmatrix} . \begin{bmatrix}
  \frac{3}{2\mk} g^{\alpha\beta}\, \partial_\mu K \partial_\nu K  & \quad 0 \\
0 &\frac{2\mk}{3}  g_{\alpha\beta} \partial_\mu K \partial_\nu K  \\
\end{bmatrix}
 . \begin{bmatrix}
\hat{\rho}_\beta{}^\nu \\
\tilde\rho^{\beta \nu} \\
\end{bmatrix}
\\ 
& = &\frac{1}{4}
\partial_\mu K \partial_\nu K  \left(  \frac{3}{2\mk}g^{\alpha\beta}  \hat{\rho}_\alpha{}^\mu \hat{\rho}_\beta{}^\nu +\frac{2\mk}{3} \, g_{\alpha\beta} \,\, \tilde\rho^{a \mu} \tilde\rho^{\beta \nu} \right)\, , 
\label{eq:D-terms-new}
\eea
with $g_{\a\b} = -\frac{3}{2\mk}{\rm Im}\, \hat{\cal K}_{\alpha\beta}$ and $g^{\a\b}$ its inverse. It is then easy to see that the full flux potential $V = V_F + V_D$ can be written of the bilinear form \eqref{VF}, by simply adding \eqref{eq:NSorbitsNew2} to the polynomials \eqref{rhos} and enlarging $Z$ accordingly. 


\section{Analysis of the potential}
\label{s:fluxpot}

While axion polynomials allow for a simple, compact expression for the flux potential, finding its vacua in full generality is still quite a formidable task. In this section we discuss some general features of this potential that, in particular, will lead to a simple Ansatz for the search of vacua. In the following section we will implement these observations for the case of compactifications with geometric fluxes. As the D-term piece of the potential will not play a significant role, in this section we will neglect its presence by considering compactifications such that $h_+^{1,1} = 0$.  Nevertheless, the whole discussion can be easily extended to a more general case.

\subsection{Stability and F-terms}\label{ss:fterms}

Given the F-term potential \eqref{eq:potential}, one may directly compute its first derivatives to find its extrema and, subsequently, its second derivatives to check their perturbative stability. However, as (meta)stability may be rather delicate to check for non-supersymmetric vacua, it is always desirable to have criteria that simplify the stability analysis. 

A simple criterium to analyse vacua metastability for F-term potentials in 4d supergravity was developed in \cite{GomezReino:2006dk,GomezReino:2006wv,GomezReino:2007qi,Covi:2008ea,Covi:2008zu}, with particular interest on de Sitter vacua. As argued in there, the sGoldstino direction in field space is the one more likely to become tachyonic in generic de Sitter vacua. Therefore, a crucial necessary condition for metastability is that such a mass is positive. Interestingly, the stability analysis along the sGoldstino direction can essentially be formulated in terms of the K\"ahler potential, which allows analysing large classes of string compactifications simultaneously. 

Following the general discussion in \cite{GomezReino:2006dk,GomezReino:2006wv,GomezReino:2007qi,Covi:2008ea,Covi:2008zu}  the sGoldstino masses can be estimated by
\be
m^2 = (3m_{3/2}^2 + \kappa_4^2 V)\, \hat{\sig}-  \frac{2}{3} \kappa_4^2 V\, ,
\label{sgoldmass}
\ee
where $m_{3/2} = e^{K/2} |W|$ is the gravitino mass, and
\be
\hat{\sig} = \frac{2}{3}-R_{A\bar B C \bar D} f^{A} f^{\bar B} f^{C} f^{\bar D}\, ,
\label{sigma}
\ee
is a function of the normalised F-terms $f_A = \frac{G_A}{(G^AG_A)^{1/2}}$ with $G_A = D_{A} W$, and the Riemann curvature tensor $R_{A\bar{B}C\bar{D}}$. Therefore, if $V$ is positive so must be $\hat{\sig}$, or else the extremum will be unstable. Reversing the logic, the larger $\hat{\sigma}$ is, the more favorable will be a class of extrema to host metastable  vacua. 

It is quite instructive to compute $\hat{\sigma}$ in our setup. Notice that because the Riemann curvature tensor only depends on the K\"ahler potential, the analysis can be done independently of which kind of fluxes are present. Moreover, because the moduli space metric factorises, $R_{A\bar B C \bar D} \neq 0$ only if all indices correspond to either K\"ahler or complex structure directions. As a consequence, the normalised F-terms can be expressed as
\be
f_A = \left({\rm cos}\, \b\, g_a, {\rm sin}\, \b\, g_\mu\right)\, 
\ee
where $g_a = \frac{G_a}{(G^aG_a)^{1/2}}$, $g_\mu  = \frac{G_\mu}{(G^\mu G_\mu)^{1/2}}$ are the normalised F-terms in the K\"ahler and complex structure sectors, respectively, and ${\rm tan}\, \b = \frac{(G^\mu G_\mu)^{1/2}}{(G^aG_a)^{1/2}}$. Therefore we have that
\be
\hat{\sig} = \frac{2}{3}-  \left({\rm cos}\, \b\right)^4 R_{a\bar b c \bar d}\, g^{a} g^{\bar b} g^{c} g^{\bar d} -  \left({\rm sin}\, \b\right)^4 R_{\mu\bar{\nu} \sigma\bar{\rho}}\, g^{\mu} g^{\bar \nu} g^{\sigma} g^{\bar \rho} \, .
\label{sigma2}
\ee
Following the discussion of Appendix \ref{ap:curvature}, one finds that the terms $R_{a\bar b c \bar d}\, g^{a} g^{\bar b} g^{c} g^{\bar d}$ and $R_{\mu\bar{\nu} \sigma\bar{\rho}}\, g^{\mu} g^{\bar \nu} g^{\sigma} g^{\bar \rho}$ are respectively minimized  by
\be
{g}_a = \frac{\g_K}{\sqrt{3}} K_a\, , \quad  {g}_\mu = \frac{\g_Q}{2} K_\mu\, ,
\label{partialmax}
\ee
where $\g_K, \g_Q \in \C$ are such that $|\g_K|^2 = |\g_Q|^2 = 1$.  In this case we have that
\be
\hat{\sigma} = \frac{2}{3} - \left({\rm cos}\, \b\right)^4 \frac{2}{3} - \left({\rm sin}\, \b\right)^4 \frac{1}{2}\, ,
\label{sigma3}
\ee
and it is positive for any value of $\b$. The choice \eqref{partialmax} corresponds to F-terms of the form
\be
G_A =\left\{G_a, G_\mu \right\}=\left\{\a_K K_a,\a_Q K_\mu \right\}\, ,
\label{solsfmax}
\ee
with  $\a_K, \a_Q \in \C$, the maximum value of \eqref{sigma3} being attained for $\a_K = \a_Q$ or equivalently $\tan\beta=2/\sqrt{3}$. Remarkably, the explicit branches of vacua obtained in \cite{Marchesano:2019hfb} have this F-term pattern.\footnote{More precisely, {\bf S1} vacua branches in \cite{Marchesano:2019hfb} are of the form \eqref{solsfmax}. The solutions found within the branches {\bf S2} correspond to cases where the complex structure metric factorises in two, and so their F-terms are specified in terms of a third constant $\a$. Finally, F-terms for Minkowski vacua with D6-brane moduli also have a similar structure, except that \eqref{solsfmax} should be written in terms of contravariant F-terms \cite{Escobar:2018tiu}.} In the following we will explore type IIA flux vacua whose  F-terms are of the form \eqref{solsfmax}, assuming that they include a significant fraction of perturbatively stable vacua. It would be interesting to extend our analysis to other possible maxima of  $\hat\sigma$ not captured by \eqref{partialmax}.

\subsubsection*{An F-term Ansatz}

As it turns out, \eqref{solsfmax} can be easily combined with the bilinear formalism used in the previous section.
Indeed, as pointed out in \cite{Herraez:2018vae}, F-terms can be easily expressed in terms of the axion polynomials $\rho_\cA$. The expressions in \cite{Herraez:2018vae} can be  generalised to the more involved flux superpotential \eqref{eq:Wgen} and \eqref{eq:WgenNS}, obtaining that
\begin{align}
G_a =&\left[\rho_a-\mk_{ab}\tr^b_\mu u^\mu-\frac{3}{2}\frac{\mk_a}{\mk}\left(t^a\rho_a+u^\mu\rho_\mu-\frac{1}{2}\mathcal{K}_b\tilde{\rho}^b_\mu u^\mu+\frac{1}{6}\mk\tr\right)\right]\nonumber\\ +&i\left[\mk_{ab}\tr^b+\rho_{a\mu}u^\mu+\frac{3}{2}\frac{\mk_a}{\mk}\left(\rho_0-t^au^\mu\rho_{a\mu}-\frac{1}{2}\mk_b\tr^b-\frac{1}{6}\mk \tilde{\rho}_\mu u^\mu\right)\right]\, \label{eq: F-Ta}\, ,\\
G_\mu =&\left[\rho_\mu-\frac{1}{2}\mk_a\tr^a_\mu+\frac{\p_\mu K}{2}\left(t^a\rho_a+u^\mu\rho_\mu-\frac{1}{2}\mathcal{K}_b\tilde{\rho}^b_\mu u^\mu-\frac{1}{6}\mk\tr\right)\right]\nonumber \\ +&i\left(t^a\rho_{a\mu}-\frac{1}{6}\mk \tilde{\rho}_\mu-\frac{\p_\mu K}{2}\left(\rho_0-t^au^\mu\rho_{a\mu}-\frac{1}{2}\mk_b\tr^b+\frac{1}{6}\mk \tilde{\rho}_\mu u^\mu\right)\right)\, .
\label{eq: F-Umu}
\end{align}

Therefore, to realise \eqref{solsfmax}, one needs to impose the following on-shell conditions
\bes
\label{proprho}
\begin{align}
   \rho_a-\mk_{ab}\tr^b_\mu u^\mu & = \ell_s^{-1} {\mathcal P}\, \partial_a K \label{eq: f-term prop rho_a}\, ,\\
       \mathcal{K}_{ab}\tilde{\rho}^b+\rho_{a\mu}u^\mu & =  \ell_s^{-1} {\mathcal Q}\, \partial_a K \label{eq: f-term prop rho^a}\, ,\\
    \rho_\mu-\frac{1}{2}\mk_a\tr^a_\mu & =  \ell_s^{-1} \cM\, \partial_\mu K\, ,\\
   t^a\rho_{a\mu}-\frac{1}{6}\mk \tilde{\rho}_\mu  & =  \ell_s^{-1}\cN\, \partial_\mu K\, , \label{eq: f-term prop rhoak}
\end{align}
\ees
where ${\mathcal P}$, ${\mathcal Q}$, $\cM$, $\cN$ are real functions of the moduli. In the next section we will impose these conditions for compactifications with geometric fluxes, obtaining a simple Ansatz for the search of type IIA flux vacua.

\subsection{Moduli and flux invariants}
\label{ss:invariants}

If instead of the above Ansatz we were to apply the more standard strategy of \cite{Marchesano:2019hfb}, we would compute the first and second derivatives of the potential \eqref{eq:potential}, to classify its different families of extrema and determine the perturbative stability of each of them. As pointed out in \cite{Herraez:2018vae} for the Calabi--Yau case, the derivatives of the axion polynomials \eqref{RRrhos} and \eqref{NSrhos} are themselves combinations of axion polynomials, see Appendix \ref{ap:conv} for the expressions in our more general setup. As a result, all the derivatives of the potential are functions of the saxions $\{t^a, u^\mu\}$ and the $\rho_\cA$, and in particular the extrema conditions $\p V|_{\rm vac} =0$ amount to algebraic equations involving both:
\be
\left(\p_{\a} V\right) (t^a, u^\mu, \rho_\cA)|_{\rm vac} = 0 \, ,
\label{extrema}
\ee
where $\a$ runs over the whole set of moduli $\{b^a, \xi^\mu, t^a, u^\mu\}$. The fact that the extrema equations depend on the quantised fluxes $q_\cA$ only through the $\rho_\cA$ is not surprising, as these are the gauge invariant quantities of the problem  \cite{Bielleman:2015ina,Carta:2016ynn}. In addition, because in our approximation the axions $\{b^a, \xi^\mu\}$ do not appear in the K\"ahler potential and in the superpotential they appear polynomially, they do not appear explicitly in \eqref{extrema}, but only through the $\rho_\cA$ as well. Therefore, finding the extrema of the F-term potential amounts to solve a number of algebraic equations on $\{t^a, u^\mu, \rho_\cA\}$. 

This simplifying picture may however give the impression that the more fluxes that are present, the less constrained the system of equations is. Indeed, \eqref{extrema} always amounts to $2 (1+ h^{1,1}_- + h^{2,1})$ equations, while the number of unknowns is $1 + h^{1,1}_- + h^{2,1} + n_q$, with $n_q$ the number of different $\rho$'s, which depends on the fluxes that we turn on. For Calabi--Yau with $p$-form fluxes $n_q = 3 + 2h^{1,1}_- + h^{2,1}$, while by including geometric and non-geometric fluxes we can increase it up to $n_q = 2 (2 + h^{2,1}) (1 + h^{1,1}_-)$. From this counting, it would naively seem that the more fluxes we have, the easier it is to solve the extrema equations. This is however the opposite of what is expected for flux compactifications. 

The solution to this apparent paradox is to realise that the $\rho_\cA$ are not fully independent variables, but are constrained by certain relations that appear at linear and quadratic order in them. Such relations turn out to be crucial to properly describe the different branches of vacua. In the following we will describe them for different cases in our setup.

\subsubsection*{Calabi--Yau with $p$-form fluxes}

Let us consider the case where only the fluxes $F_{2n}$, $H$ are turned on, while $f = Q = R = 0$. The moduli stabilisation analysis reduces to that in \cite{Marchesano:2019hfb}, and the extrema conditions reduce to $2 h^{1,1}_- + h^{2,1} + 2$ because only one linear combination $h_\mu\xi^\mu$ of complex structure axions appears in the F-term potential. In this case the vector of axion polynomials $\rho_\cA=(\rho_0,\rho_a,\tilde{\rho}^a,\tilde{\rho},\rho_\mu)$ has $3 + 2h^{1,1}_- + h^{2,1}$ entries, but several are independent of the axions. Indeed, at the linear level 
\be 
 \tilde{\rho}=\ell_s ^{-1} m\, ,\qquad \quad    \rho_\mu=  \ell_s^{-1} h_\mu\, ,
 \label{invCYl}
\ee
are axion-independent, while at the quadratic level
\be
 \tilde{\rho}\rho_{a}-\frac{1}{2}\mathcal{K}_{abc}\tilde{\rho}^b\tilde{\rho}^c\, =\, \ell_s^{-2}\left(m e_a -  \frac{1}{2}\mathcal{K}_{abc} m^bm^c\right)\, ,
  \label{invCYq}
\ee
is also independent of the axions. If we fix the flux quanta $q_\cA = (e_0,  e_b,   m^b,  m,  h_\mu)$, the value of \eqref{invCYl} and \eqref{invCYq} will be fixed, and $\rho_\cA$ will take values in a $(1 + h^{1,1}_-)$-dimensional orbit. This orbit corresponds to the number of axions that enter the F-term potential, and so taking these constraints into account allows to see \eqref{extrema} as a determined system. 

Interestingly, the quadratic invariant \eqref{invCYq} was already identified in \cite{DeWolfe:2005uu} as the quantity that determines the value of the K\"ahler saxions in supersymmetric vacua of this kind. In fact, this is also true for non-supersymmetric vacua \cite{Marchesano:2019hfb}. One has that
\be
m e_a -  \frac{1}{2}\mathcal{K}_{abc} m^bm^c = \tilde{A} \CK_a\, ,
\ee
with $\tilde{A} \in \mathbb{R}$ fixed for each branch of vacua. Moreover, for the branches satisfying \eqref{solsfmax}, the complex structure saxions are fixed in terms of the fluxes as $h_\mu = \hat{A} \cK \p_\mu K$, with $\hat{A}$ constant. Therefore the fluxes fix both the saxions and the allowed orbit for the $\rho_{\cA}$. Finding the latter in terms of \eqref{extrema} is equivalent to finding the values of $b^a$ and $h_\mu\xi^\mu$.

\subsubsection*{Adding geometric fluxes}

Let us now turn to compactifications with fluxes $F_{2n}$, $H$, $f$, while keeping $Q = R = 0$. The number of axions $\xi^\mu$ that enter the scalar potential now corresponds to the dimension of the vector space spanned by $\langle h_\mu, f_{a\mu}\rangle$, for all possible values of $a$. If we see $f_{a\mu}$ as a $h^{1,1}_-  \times (h^{2,1}+1)$ matrix of rank $r_f$, the number of relevant entries on $\rho_\cA=(\rho_0,\rho_a,\tilde{\rho}^a,\tilde{\rho},\rho_\mu, \rho_{a\mu})$ is $2 + (2+ r_f)h^{1,1}_- + (1+ r_f)(1 +h^{2,1})- r_f^2$. At the linear level the invariants are
\be 
 \tilde{\rho}=\ell_s ^{-1} m\, ,\qquad \quad    \rho_{a\mu}=  \ell_s^{-1} f_{a\mu}\, ,
 \label{invmetl}
\ee
while at the quadratic level we have
\be
 \tilde{\rho}\rho_\mu-\tilde{\rho}^a\rho_{a\mu} = \ell_s^{-2} \left(mh_\mu - m^a f_{a\mu}\right)  \, , \qquad  c^{a} \left(\tilde{\rho}\rho_{a}-\frac{1}{2}\mathcal{K}_{\bar{abc}}\tilde{\rho}^b\tilde{\rho}^c\right)\, .
 \label{invmetq}
\ee
Here the $c^a \in \mathbb{Z}$ are such that $c^a \rho_{a\mu} = 0$ $\forall \mu$, so there are $h^{1,1}_- - r_f$ of this last class of invariants. Taking all these invariants into account we find that $\rho_\cA$ takes values in a $(1 + h^{1,1}_- + r_f)$-dimensional orbit,\footnote{If $d^a f_{a\mu} = h_\mu$ for some $d^a \in \R$, then the $\rho_\cA$ draw a  $(h^{1,1}_- + r_f)$-dimensional orbit, and one less axion is stabilised. As a result one can define an additional flux invariant.  See next section for an example.} signalling the number of stabilised axions. In other words, with the inclusion of metric fluxes the orbit of allowed $\rho_\cA$ increases its dimension, which implies that more moduli, in particular more axions $\xi^\mu$ are fixed by the potential. As in the CY case, the saxions are expected to be determined in terms of these invariants.

\subsubsection*{Adding non-geometric fluxes}

The same kind of pattern occurs when non-geometric fluxes are included. If one sets $R = 0$, the invariants at the linear level are $\tilde{\rho}$ and $\tilde{\rho}_\mu^a$, as well the combinations $c^a d^\mu \rho_{a\mu}$ with $c^a, d^\mu \in \mathbb{Z}$ such that $c^a d^\mu \cK_{abc}Q^c_\mu = 0$, $\forall b$. At the quadratic level, the first invariant in \eqref{invmetq} is replaced by 
\be
\tilde{\rho}\rho_\mu-\tilde{\rho}^a\rho_{a\mu} + \rho_a \tilde{\rho}^a_\mu \, ,
   \label{invngq}
\ee
where we have taken into account the Bianchi identity $f_{a[\mu}\, Q^a{}_{\nu]} = 0$.
Additionally, the second invariant  in \eqref{invmetq} may also survive if there are choices of $c^a \in \mathbb{Z}$ such that $c^a \rho_{a\mu} \xi^\mu = 0$ $\forall \xi^\mu$.
Finally, when all kind of fluxes are nonvanishing, the only invariant at the linear level is $R_\mu$, and some particular choices of $\tilde{\rho}^a_\mu$ and $\rho_{a\mu}$. At the quadratic level we have the generalisation of \eqref{invngq}
\be
\tilde{\rho}\rho_\mu-\tilde{\rho}^a\rho_{a\mu} + \rho_a \tilde{\rho}^a_\mu -\rho_0\tilde{\rho}_\mu\, ,
   \label{invngr}
\ee
where we have imposed the Bianchi identity $\rho_{[\mu} \, \tilde{\rho}_{\nu]} - \rho_{a[\mu}\, \tilde{\rho}^a{}_{\nu]} = h_{[\mu} \, R_{\nu]} - f_{a[\mu}\, Q^a{}_{\nu]} = 0$, see Appendix \ref{ap:conv}. Notice that this invariant and its simpler versions are nothing but  the D6-brane tadpole \eqref{DFtadpole} induced by fluxes. We also have the new invariants
\be
\tilde{\rho}_{[\mu}^a \tilde{\rho}_{\nu]}\, ,\qquad \qquad   \rho_{a(\mu}\tilde{\rho}_{\nu)}-\mathcal{K}_{abc}\tilde{\rho}^b_\mu\tilde{\rho}^c_\nu\, ,
\label{finalNGinv}
\ee
where as above $(\ )$ and $[\ ]$ stand for symmetrisation and anti-symmetrisation of indices, respectively. Finally, if the second invariant in \eqref{finalNGinv} vanishes, or in other words if we have $f_{a(\mu}Q_{\nu)}=\mathcal{K}_{abc}Q^b_\mu Q^c_\nu$, then
\be
\rho_{a(\mu}\tilde{\rho}_{\nu)}^a - 3 \rho_{(\mu} \tilde{\rho}_{\nu)}\, ,
\label{ffinalNGinv}
\ee
is also an invariant.\footnote{Remarkably, both \eqref{ffinalNGinv} and the second invariant in \eqref{finalNGinv} vanish if the ``missing" Bianchi identities $f_{a(\mu}Q_{\nu)}=\mathcal{K}_{abc}Q^b_\mu Q^c_\nu$ and $f_{a(\mu}Q_{\nu)}^a = 3 h_{(\mu} R_{\nu)}$ proposed in \cite{Gao:2018ayp} turn out to hold generally.}


\section{Geometric flux vacua}
\label{s:geovacua}

In this section we would like to apply our previous results to the search of vacua in type IIA flux compactifications. For concreteness, we focus on those configurations with $p$-form and geometric fluxes only, leaving the systematic search of non-geometric flux vacua for future work. As we will see, for geometric flux vacua the Ansatz formulated in the last section, which amounts to impose on-shell F-terms of the form \eqref{solsfmax},  
forbids de Sitter solutions. In contrast, we find two branches of AdS extrema corresponding to our Ansatz, one supersymmetric and one non-supersymmetric. The perturbative stability of the latter will be analysed in the next section. 

\subsection{The geometric flux potential}

Let us first of all summarise our previous results and restrict them to the case of $p$-form and geometric fluxes. The scalar potential reads $V = V_F + V_D$, with
\begin{align}
   \kappa_4^2 V_F =\, &e^K\left[4\rho_0^2+g^{ab}\rho_a\rho_b+\frac{4\mathcal{K}^2}{9}g_{ab}\tilde{\rho}^a\tilde{\rho}^b+\frac{\mathcal{K}^2}{9}\tilde{\rho}^2\right.\nonumber\\
    + &\left.c^{\mu\nu}\rho_\mu\rho_\nu+\left(\tilde{c}^{\mu\nu}t^at^b+g^{ab}u^\mu u^\nu \right)\rho_{a\mu}\rho_{b\nu}-\frac{4\mathcal{K}}{3}u^\nu \tilde{\rho}^a\rho_{a\nu}+\frac{4\mathcal{K}}{3}u^\nu \tilde{\rho}\rho_\nu\right]\, ,
    \label{eq:potentialgeom}\\
\kappa_4^2 V_D = \, &\frac{3}{8\mathcal{K}} \partial_\mu K \partial_\nu K \,  g^{\alpha\beta} \, \hat{\rho}_\alpha{}^\mu \hat{\rho}_\beta{}^\nu\, .\label{eq: D-potentialgeom}
\end{align}
The definitions for $g^{ab}$,  ${c}^{\mu\nu}$, $\tilde{c}^{\mu\nu}$ and $g^{\alpha\beta}$ are just as in section \ref{s:IIAorientifold}, while the $\rho_\cA$ simplify to
\bes
\label{RRrhosgeom}
\begin{align}
  \ell_s  \rho_0&=e_0+e_ab^a+\frac{1}{2}\mathcal{K}_{abc}m^ab^bb^c+\frac{m}{6}\mathcal{K}_{abc}b^ab^bb^c+\rho_\mu\xi^\mu\, , \label{eq:rho0g}\\
 \ell_s   \rho_a&=e_a+\mathcal{K}_{abc}m^bb^c+\frac{m}{2}\mathcal{K}_{abc}b^bb^c+\rho_{a\mu}\xi^\mu \, ,  \label{eq:rho_ag}\\
  \ell_s  \tilde{\rho}^a&=m^a+m b^a \, ,  \label{eq:rho^ag}\\
 \ell_s   \tilde{\rho}&=m \, ,   \label{eq:rhomg}\\
 \ell_s    \rho_\mu&=h_\mu+f_{a\mu}b^a \, , \\
 \ell_s   \rho_{a\mu}&=f_{a\mu} \, ,  \label{eq: ho_ak} \\
 \ell_s   \hat{\rho}_{\a}^{\mu}&=\hat{f}_{\a}^{\mu} \, . 
\end{align}   
\ees 

Using these explicit expressions one may compute the first order derivatives of the scalar potential with respect to the axions $\{\xi^\mu, b^a\}$ and saxions $\{u^\mu, t^a\}$ of the compactification. As expected the extrema conditions are of the form \eqref{extrema}, with:

\vspace*{.5cm}

\textbf{Axionic directions}

\bes
\label{paxions}
\begin{equation}
\label{paxioncpx}
  e^{-K}\frac{\partial V}{\partial \xi^\mu}=8\rho_0\rho_\mu +2g^{ab}\rho_a\rho_{b\mu} \, ,
\end{equation}
\begin{equation}
\label{paxionk}
 e^{-K}\frac{\partial V}{\partial b^a}= \ 8\rho_0\rho_a+\frac{8}{9}\mathcal{K}^2g_{ac}\tilde{\rho}\tilde{\rho}^c  +2\mathcal{K}_{abd}g^{bc}\rho_c\tilde{\rho}^d+2c^{\mu\nu}\rho_{a\mu}\rho_\nu\, ,
\end{equation}
\ees

\vspace*{.5cm}

\textbf{Saxionic directions}

\bes
\label{psaxions}
\begin{eqnarray}
\label{psaxioncpx}
 e^{-K}  \frac{\partial V}{\partial u^\mu} & =  & e^{-K} V_F\partial_\mu K+\frac{4}{3}\mathcal{K}\tilde{\rho} \rho_\mu +\partial_\mu c^{\kappa\sigma}\rho_\kappa\rho_\sigma - \frac{4}{3}\mathcal{K}\tilde{\rho}^a\rho_{a\mu} +2g^{ab}\rho_{a\mu}\rho_{b\nu}u^\nu\\ \nonumber
    & & + t^at^b(\partial_\mu c^{\kappa\sigma}\rho_{a\kappa}\rho_{b\sigma}-8\rho_{a\mu}\rho_{b\nu}u^\nu )
    + \frac{3}{4\mathcal{K}}e^{-K}\partial_\mu \partial_\sigma K \partial_\nu K \,  g^{\alpha\beta} \, \hat{\rho}_\alpha{}^\sigma \hat{\rho}_\beta{}^\nu\, ,
\end{eqnarray}
\begin{eqnarray}
\label{psaxionk}\nonumber
e^{-K} \frac{\partial V}{\partial t^a} &= & e^{-K} V_F\partial_{a}K+\partial_{a}\left(\frac{4}{9}\mathcal{K}^2\tilde{\rho}^b\tilde{\rho}^c
 g_{bc}\right)+\partial_{a}g^{cd}\rho_c\rho_d+\mathcal{K}_a\tilde{\rho}\left(\frac{2}{3}\mathcal{K}\tilde{\rho}+4u^\mu{\rho}_\mu \right)\\
 & & - 4 \mathcal{K}_a \tilde{\rho}^b\rho_{b\nu}u^\nu +2\tilde{c}^{\mu\nu}t^c\rho_{a\mu}\rho_{c\nu}+\partial_a g^{bc}\rho_{b\mu}u^\mu \rho_{c\nu}u^\nu\nonumber\\  & & +\frac{3}{8\mathcal{K}}e^{-K} \partial_\mu K \partial_\nu K \,  \p_a g^{\alpha\beta} \, \hat{\rho}_\alpha{}^\mu \hat{\rho}_\beta{}^\nu\,-\frac{9\mathcal{K}_a}{8\mathcal{K}^2}e^{-K} \partial_\mu K \partial_\nu K \,   g^{\alpha\beta} \, \hat{\rho}_\alpha{}^\mu\hat{\rho}_\beta^\nu\, .
\end{eqnarray}
\ees

\subsection{de Sitter no-go results revisited}
\label{subsec: no-go's}

From \eqref{psaxions} one can obtain the following off-shell relation
\begin{align}
\nonumber
    & u^\mu \partial_{u^\mu} V +x \, t^a \partial_{t^a} V=- (4+3x)V_F - (2+x) V_D + 4e^K\left[x\left(\frac{1}{2}g^{bc}\rho_b\rho_c+\frac{4\mathcal{K}^2}{9}g_{bc}\tilde{\rho}^b \tilde{\rho}^c +\frac{\mathcal{K}^2}{6}\tilde{\rho}^2\right) \right. \\
    & +\left. \frac{1}{2}c^{\mu\nu}\rho_\mu\rho_\nu  +\left(\frac{1}{3}+x\right)\mathcal{K}u^\nu \left(\tilde{\rho}\rho_\nu -\tilde{\rho}^b\rho_{b\nu}\right)  +\frac{1}{2}(1+x)(\tilde{c}^{\mu\nu}t^bt^c+g^{bc}u^\mu u^\nu )\rho_{b\mu}\rho_{c\nu}\right]\, ,
   \label{eq: arbitrary combination of partial derivatives}
\end{align}
with $x \in \R$ an arbitrary parameter. Different choices of $x$ will lead to different equalities by which one may try to constrain the presence of extrema with positive energy, in the spirit of  \cite{Hertzberg:2007wc,Flauger:2008ad}. In practice it is useful to rewrite this relation as
\begin{equation}
    u^\mu \partial_{u^\mu} V +x t^a \partial_{t^a} V = - 3V + \Xi_x\, ,
\end{equation}
where, for instance, the choice $x=1/3$ leads to 
\begin{equation}
    \label{eq: cosmo const relation}
   \Xi_{1/3} = \frac{2}{3} V_D  +  4e^K\left[-2\rho_0^2-\frac{1}{3}g^{bc}\rho_b\rho_c-\frac{2}{27}\tilde{\rho}^b\tilde{\rho}^c\mathcal{K}^2g_ {bc}+\frac{1}{6}(t^at^b\tilde{c}^{\mu\nu}+g^{ab}u^\mu u^\nu )\rho_{a\mu}\rho_{a\nu}\right] ,
\end{equation}
while the choice $x=1$ gives
\begin{equation}
 \Xi_1 = 4e^K\left[\frac{\mathcal{K}^2}{18}\tilde{\rho}^2-4\rho_0^2-\frac{1}{2}g^{ab}\rho_a\rho_b-\frac{1}{2}c^{\mu\nu}\rho_\mu\rho_\nu\right]\, .
    \label{eq: cosmo cons relation 2}
\end{equation}
Extrema of positive energy require $\p V =0$ and $V >0$, and so necessarily both \eqref{eq: cosmo const relation} and \eqref{eq: cosmo cons relation 2} should be positive. It is easy to see that this requires that both the Romans' parameter $\tilde{\rho}$ and geometric fluxes (either $\rho_{a\mu}$ or $\hat{\rho}_\alpha^\mu$)  are present, in agreement with previous results in the literature \cite{Haque:2008jz,Caviezel:2008tf,Flauger:2008ad,Danielsson:2009ff,Danielsson:2010bc,Danielsson:2011au}. In that case, it is unlikely that the potential satisfies an off-shell inequality of the form proposed in \cite{Obied:2018sgi}, at least at the classical level. 

In our formulation one can make more precise which kind of fluxes are necessary to attain de Sitter extrema. For this, let us express the last term of \eqref{eq: cosmo const relation} as
\be
(t^at^bc^{\mu\nu}+g^{ab}u^\mu u^\nu -4t^at^bu^\mu u^\nu )\rho_{a\mu}\rho_{a\nu} = \left[t^at^bc_{\rm P}^{\mu\nu}+u^\mu u^\nu g_{\rm P}^{ab} -\frac{5}{3} t^at^bu^\mu u^\nu \right]\rho_{a\mu}\rho_{a\nu}\, ,
\label{ccgeomterm}
\ee
where $g_{\rm P}^{ab}$, $c_{\rm P}^{\mu\nu}$ are the primitive components of the K\"ahler and complex structure metric, respectively. That is 
\be
\label{eq: primitive metric}
g_{\rm P}^{ab} = \frac{2}{3}\left(t^at^b - \mathcal{K}\mathcal{K}^{ab}\right)\, , \qquad \qquad c_{\rm P}^{\mu\nu} = \frac{1}{3}u^\mu u^\nu - 4G_Q G_Q^{\mu\nu}\, ,
\ee
where $G_Q = e^{-K_Q}$ and $G_Q^{\mu\nu}$ is the inverse of $\p_\mu \p_\nu G_Q$. These metric components have the property that they project out the K\"ahler potential derivatives along the overall volume and dilaton directions, namely $g_{\rm P}^{ab} \p_b K = c_{\rm P}^{\mu\nu} \p_\nu K =0$. So in order for the bracket in \eqref{eq: cosmo const relation} to be positive, the geometric fluxes $\rho_{a\mu}$ not only must be non-vanishing, but they must also be such that
\be
t^a \rho_{a\mu}\, t^b \rho_{a\nu}\, c_{\rm P}^{\mu\nu}+  \rho_{a\mu} u^\mu \, \rho_{a\nu} u^\nu\, g_{\rm P}^{ab} \neq 0\, .
\label{primcond}
\ee
In other words, either the vector $\rho_{a\mu} u^\mu$ is not proportional to $\p_a K$ or the vector $t^a \rho_{a\mu}$ is not proportional to $\p_\nu K$. The condition is likely to be satisfied at some point in field space, but in order to allow for a de Sitter extremum it must be satisfied on-shell as well. 

Remarkably, we find that the F-term Ansatz of section \ref{ss:fterms} forbids de Sitter extrema. Indeed, if we impose that the on-shell relations \eqref{proprho} are satisfied with the non-geometric fluxes turned off (cf. \eqref{proprhog} below) we obtain that, on-shell
\be
t^a \rho_{a\mu}\, t^b \rho_{a\nu}\, c_{\rm P}^{\mu\nu}+  \rho_{a\mu} u^\mu \, \rho_{a\nu} u^\nu\, g_{\rm P}^{ab} = \frac{4}{9} \cK^2 g^{\rm P}_{ab} \tilde{\rho}^a \tilde{\rho}^b \, ,
\label{primvalue}
\ee
with $g^{\rm P}_{ab}$ the inverse of $g_{\rm P}^{ab}$ in the primitive sector. Even if this term is positive, it can never be bigger than the other negative contributions within the bracket in \eqref{eq: cosmo const relation}. In fact, after plugging \eqref{primvalue} in \eqref{eq: cosmo const relation} there is a partial cancellation between the third and fourth term of the bracket, that then becomes semidefinite negative:
\be
4e^K\left[-2\rho_0^2-\frac{1}{3}g^{ab}\rho_a\rho_b-\frac{2}{27}\tilde{\rho}^a\tilde{\rho}^b\mathcal{K}^2g_ {ab}^{\rm NP}-\frac{5}{18}t^at^bu^\mu u^\nu \rho_{a\mu}\rho_{b\nu}\right]\, ,
\ee
with $g_ {ab}^{\rm NP} = g_{ab} -  g^{\rm P}_{ab} = \frac{3}{4}\frac{\cK_a \cK_b}{\cK^2} $ the non-primitive component of the K\"ahler moduli metric.

Even if the bracket in \eqref{eq: cosmo const relation} is definite negative,  there is still the contribution from the piece $\frac{2}{3}V_D$, which is positive semidefinite. However, one can see that with the Ansatz \eqref{solsfmax} this contribution vanishes. Indeed, using the Bianchi identity $f_{a\mu} \hat{f}_\alpha{}^\mu = 0$ and \eqref{eq: f-term prop rhoak geom}, or alternatively $h_\mu \hat{f}_\alpha{}^\mu = 0$ and \eqref{eq: f-term prop rhomu geom}. one can see that the D-term $D_\alpha =\frac{1}{2} \partial_\mu K \, \hat f_\alpha{}^\mu$ vanishes, and so does $V_D$.

To sum up, for type IIA geometric flux configurations, in any region of field space in which the F-terms are of the form \eqref{solsfmax} we have that the F-term potential satisfies
\be
u^\mu \partial_{u^\mu} V+\frac{1}{3}t^a \partial_{t^a} V \leq -3V\, , 
\label{eq: no-go geom inequality}
\ee
and so de Sitter extrema are excluded. In other words:
\begin{center}
{\em In type IIA geometric flux compactifications, classical  de Sitter extrema \\  are incompatible with F-terms of the form \eqref{solsfmax}.}
\end{center}
In section \ref{sec:10d} we will interpret this result from a geometrical viewpoint. 
It would be interesting to extend this discussion to non-geometric flux compactifications, along the lines of \cite{deCarlos:2009fq,Shukla:2019dqd}, to see if this result applies there as well.

\subsection{Imposing the Ansatz}\label{imposing}

Besides the cosmological constant sign, let us see other constraints that the on-shell condition \eqref{solsfmax} leads to. By switching off all non-geometric fluxes, \eqref{proprho} simplifies to
\bes
\label{proprhog}
\begin{align}
   \rho_a & = \ell_s^{-1} {\mathcal P}\, \partial_a K \label{eq: f-term prop rho_a geom}\, ,\\
       \mathcal{K}_{ab}\tilde{\rho}^b+\rho_{a\mu}u^\mu & =  \ell_s^{-1} {\mathcal Q}\, \partial_a K \label{eq: f-term prop rho^a geom}\, ,\\
    \rho_\mu & =  \ell_s^{-1}\cM\, \partial_\mu K\, ,  \label{eq: f-term prop rhomu geom} \\
   t^a\rho_{a\mu } & =  \ell_s^{-1} \cN\, \partial_\mu K\, , \label{eq: f-term prop rhoak geom}
\end{align}
\ees
where again ${\mathcal P}$, ${\mathcal Q}$, $\cM$, $\cN$ are real functions of the moduli. Such functions and other aspects of this Ansatz are constrained by the extrema conditions \eqref{paxions} and \eqref{psaxions} with which they must be compatible. Indeed, plugging \eqref{proprhog} into \eqref{paxions} one obtains  
\bes
\label{paxionsA}
\begin{equation}
\label{paxioncpxA}
8 \left(\rho_0\cM -  {\mathcal P}\cN\right) \partial_\mu K = 0 \, ,
\end{equation}
\begin{equation}
\label{paxionkA}
\left[ 8  {\mathcal P} (\rho_0 -  {\mathcal Q}) - \frac{1}{3} \tilde{\rho}  \cK \left(-2 {\mathcal Q} + 8 \cN \right)   \right]  \partial_a K   + \left[ \frac{4}{3} \CK\tilde{\rho} + 8  {\mathcal P} - 8 \cM \right] \rho_{a\mu} u^{\mu} = 0 \, ,
\end{equation}
\ees
which must be satisfied on-shell. Even when both brackets in \eqref{paxionkA} vanish, this equation implies that on-shell
\be
\rho_{a\mu} u^{\mu} \propto \p_a K\, , \qquad {\rm and} \qquad \tilde{\rho}^a \propto t^a\, ,
\ee
simplifying the Ansatz. More precisely, we are led to the following on-shell relations
\bes
\label{Ansatz}
\begin{align}
   \ell_s \rho_0&= A \CK\, , \label{eq: ans rho0}\\
    \ell_s\rho_a&= B \cK \partial_a K  \, ,  \label{eq: ans rho_a}\\
    \ell_s\tilde{\rho}^a&= C t^a  \, ,  \label{eq: ans rho^a}\\
    \ell_s\tilde{\rho}&= D \, ,  \label{eq: ans rhotilde}\\
    \ell_s\rho_\mu&=E\cK \partial_\mu K  \, ,   \label{eq: nsns}\\
    \ell_s\rho_{a\mu} t^a &= \frac{F}{4}  \cK\partial_\mu K  \, ,  \label{eq: geoma}\\
     \ell_s\rho_{a\mu} u^\mu &= \frac{F}{3}  \CK \partial_a K   \, , \label{eq: geomu}
\end{align}
\ees
where $A, B, C, D, E, F$ are functions of the saxions. We have extracted a factor of $\CK$ in some of them so that the expression for the on-shell equations simplifies. In terms of \eqref{Ansatz} we have that the vanishing of \eqref{paxions} amounts to
\bes
\label{paxionsAA}
\begin{equation}
\label{paxioncpxAA}
 4 AE -  BF =  0 \, ,
\end{equation}
\begin{equation}
\label{paxionkAA}
 3AB  - \frac{1}{12} CD  + B C  -  EF = 0 \, ,
\end{equation}
\ees
assuming that at each vacuum $\p_\mu K \neq 0 \neq \p_a K$. Similarly, the vanishing of \eqref{psaxions} implies
\bes
\label{psaxionsAA}
\begin{equation}
\label{psaxioncpxAA}
 4A^2 + 12B^2 +\frac{1}{3} C^2 + \frac{1}{9}D^2 + 8 E^2 -\frac{5}{6} F^2 +CF -4DE = 0\, ,
\end{equation}
\begin{eqnarray}
\label{psaxionkAA}
4A^2 + 4 B^2 - \frac{1}{9} C^2 - \frac{1}{9}D^2 + 16E^2 -\frac{5}{9}F^2  = 0\, ,
\end{eqnarray}
\ees
where we have used the identities in \cite[Appendix A]{Marchesano:2019hfb}.

Expressing the extrema equations in terms of the Ansatz \eqref{Ansatz} has the advantage that we recover a system of algebraic equations. Nevertheless, eqs.\eqref{paxionsAA} and \eqref{psaxionsAA} may give the wrong impression that we have an underdetermined system, with four equations and six unknowns $A, B, C, D, E, F$. Notice, however, that these unknowns are not all independent, and that relations among them arise when the flux quanta are fixed. Indeed, let us first consider the case without geometric fluxes, which sets $F=0$. In this case, AdS vacua require that the Roman's parameter $m$ is non-vanishing so we may assume that $D \neq 0$. Because the LHS of \eqref{paxionsAA} and \eqref{psaxionsAA}  are  homogeneous polynomials of degree two, we may divide each of them by $D^2$ to obtain four equations on four variables: $A_D = A/D$, $B_D = B/D$, $C_D = C/D$, $E_D = E/D$. The solutions correspond to $A_D= 0$ and several rational values for $B_D, C_D, E_D$, which reproduce the different {\bf S1} branches found in \cite{Marchesano:2019hfb}.\footnote{To compare to \cite{Marchesano:2019hfb} one needs to use the  dictionary: $B_D = - C_{\rm MQ}/3$, $C_D = B_{\rm MQ}$, $E_D = A_{\rm MQ}$.} Finally, the variable $D = m$ is fixed when the flux quanta are specified. 

The analysis is slightly more involved in the presence of geometric fluxes. Now we may assume that $F \neq 0$, since otherwise we are back to the previous case. Our Ansatz implies that the first flux invariant in \eqref{invmetq} is a linear combination of the vectors $(f_a)_\mu = f_{a\mu}$, as
\be
m\hat{h}_\mu \equiv m h_\mu - m^a f_{a\mu} = \left(DE - \frac{CF}{4}\right)  \cK \p_\mu K  =   \left( \frac{4DE}{F}  - C \right)  t^a  f_{a\mu} \, ,
\label{hath}
\ee
where $\cK$, $ \p_\mu K$, $ t^a$ correspond to the value of the K\"ahler saxions in the corresponding extremum, etc.  One can write the above relation as
\be
m\hat{h}_\mu = d^a f_{a\mu}\, ,
\label{condhhat}
\ee
where the constants $d^a$ are fixed once that we specify the fluxes $m$, $h_\mu$, $m^a$, $f_{a\mu}$. As a consequence, the number of stabilised complex structure axions $\xi^\mu$ is $r_f = {\rm rank}\, f_{a\mu}$, while the rest may participate in St\"uckelberg mechanisms triggered by the presence of D6-branes \cite{Camara:2005dc}.\footnote{Microscopically, \eqref{condhhat} means that $h_\mu$ is in the image of the matrix of geometric fluxes $f_{a\mu}$, and as such it is cohomologically trivial. Macroscopically, it means that the number of independent complex structure axions entering the scalar potential are $ {\rm dim} \langle h_\mu, f_{1\mu}, f_{2\mu}, \dots \rangle= {\rm rank} f_{a\mu} \equiv r_f $, and not $r_f+1$.} Strictly speaking, $d^a$ is only fixed up to an element in the kernel of $f_{a\mu}$, but this is irrelevant for our purposes. Indeed, notice that due to our Ansatz
\bea\nonumber
m\hat{e}_a \equiv  me_a -  \frac{1}{2}\mathcal{K}_{abc} m^bm^c  &= & \left(BD + \frac{C^2}{6} \right)  \cK \p_a K   - m f_{a\mu} \xi^\mu \\ 
 & = & \left[\left(\frac{3BD}{F} + \frac{C^2}{2F} \right)  u^\mu   -  D \xi^\mu  \right] f_{a\mu}  \, ,
 \label{hatea}
\eea
where again $\cK$, $u^\mu$, $\xi^\mu$ stand for the vevs at each extremum. This implies several things. First, the second set of invariants in \eqref{invmetq} vanishes identically. Second, the combination $md^a \hat{e}_a$ is fully specified by the flux quanta, without any ambiguity.  Finally in terms of 
\begin{equation} 
m^2 \hat{e}_0 \equiv m^2 e_0 - m m^a e_a  + \frac{1}{3} \mathcal{K}_{abc}m^am^bm^c  \, ,
\label{hate0}
\end{equation} 
we can define the following cubic flux invariant
\be
m^2 \hat{e}_0 -  m d^a  \hat{e}_a  =  \cK  \left[AD^2  + 3BCD + \frac{C^3}{3} + \left( \frac{4DE}{F}  - C\right) \left(3BD + \frac{C^2}{2} \right)  \right] \, .
\label{extrainv}
\ee
The existence of this additional invariant is expected from the discussion of section \ref{ss:invariants}. As we now show, $\cK$ is fixed at each extremum by the choice of the flux quanta and the Ansatz' variables. Therefore \eqref{extrainv} and $D=m$ provide two extra constraints on these variables, which together with \eqref{paxionsAA} and \eqref{psaxionsAA}  yield a determined system of algebraic equations. 

To show how  $\cK$ is specified, let us first see how the saxionic moduli are determined. First \eqref{hath} determines $(4DE-CF) \cK  \partial_\mu K $ in terms of the flux quanta, which is equivalent to determine $(4DE-CF)^{-1}  u^\mu/\cK$. Plugging this value into \eqref{eq: geomu} one fixes $(4DE/F-C)^{-1}  \partial_a K$ in terms of the fluxes, which is equivalent to fix $(4DE/F-C)  t^a$. Therefore at each extremum we have that
\be
\left(\frac{4DE}{F}-C\right)^3  \cK  \, ,
\label{vevK}
\ee
is specified by the flux quanta. Notice that this is compatible with \eqref{hath}, and we can actually use this result to fix the definition of $d^a$, by equating \eqref{vevK} with $\cK_{abc}d^ad^bd^c$.

\subsection{Branches of vacua}\label{branchvacu}

\begin{center}    
\begin{figure}[H]
    \centering
    \includegraphics[scale=1]{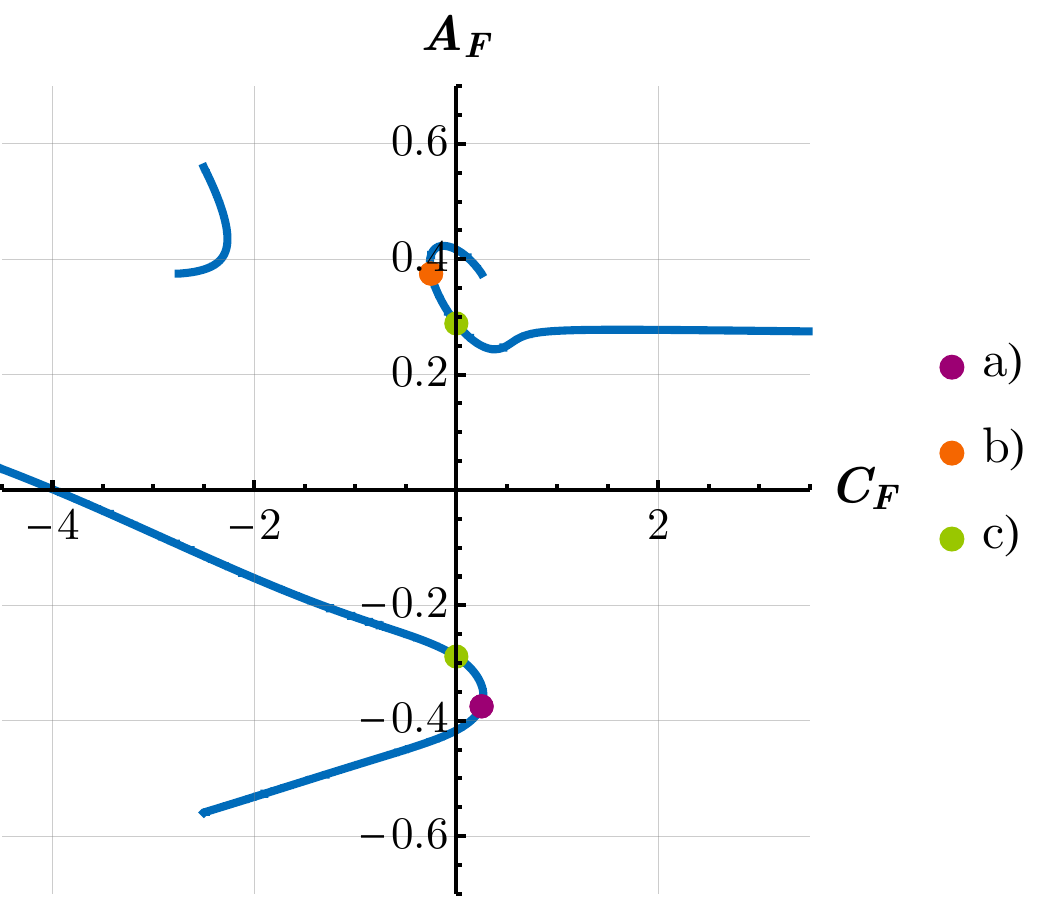}
    \caption{Set of points that verify \eqref{eq: megaeqF} (blue curve) and have $E_F^2\geq 0$. The coloured dots correspond to the particular solutions \eqref{delta=0sols}. Both curves tend asymptotically to $A_F=1/4$ for $C_F\rightarrow\pm \infty$. }
    \label{fig: generalsol}
\end{figure}
\end{center}

Let us analyse the different solutions to the algebraic equations \eqref{paxionsAA} and \eqref{psaxionsAA}. 
Following the strategy of the previous subsection, we assume that $F \neq 0$ and define $A_F = A/F$, $B_F = B/F$, $C_F = C/F$, $D_F = D/F$, $E_F = E/F$. Then, from \eqref{paxioncpxAA} we obtain
\begin{equation}
    B_F=4A_FE_F\, ,
    \label{eq: B_F}
\end{equation}
which substituted into \eqref{paxionkAA} gives the following relation
\begin{equation}
  C_FD_F= 12 E_F(12A_F^2+4A_FC_F-1)\, .
    \label{eq: C_FD_F}
\end{equation}
Then, multiplying \eqref{psaxionkAA} by $C_F^2$ and using \eqref{eq: C_FD_F} we obtain
\be
144 E_F^2 \Delta_F = C_F^2\left[36 A_F^2 - C_F^2 - 5 \right]  \, ,
  \label{eq: E_F Delta_F}
\ee
where
\be
\Delta_F =(12A_F^2+4A_FC_F-1)^2 - 4 A_F^2 C_F^2 - C_F^2\, .
   \label{eq: DeltaF def}
\ee
We have two possibilities, depending on whether $\Delta_F = 0$ or not. Let us consider both:

\begin{itemize}

\item $\Delta_F = 0$

In this case, from \eqref{eq: E_F Delta_F} and \eqref{eq: DeltaF def}, we find four different real solutions for $(A_F, C_F)$:
\bes
\label{delta=0sols}
   \begin{align}
   \label{SUSYsol}
            A_F=-\frac{3}{8}\, , \ \ \ & \ \ \ C_F=\frac{1}{4}\, ,\\
            A_F=\frac{3}{8}\, , \ \ \ & \ \ \ C_F=-\frac{1}{4}\, ,
            \label{nonSUSYdelta=0}\\
            A_F=\pm \frac{1}{2\sqrt{3}}\, , \ \ \ & \ \  \ C_F= 0 \, .
            \label{nonSUSYdelta=0C=0}
        \end{align}
\ees
Given the solution \eqref{SUSYsol}, one can solve for $D_F$ in \eqref{eq: C_FD_F} and check that \eqref{psaxioncpxAA} and \eqref{psaxionkAA} are automatically satisfied. We then find that:
\be
\eqref{SUSYsol} \ \raw \ B_F = -\frac{3}{2} E_F\, , \qquad D_F = 15E_F\, ,
\label{SUSYbranch}
\ee
with $E_F$ unfixed. Thus, at this level $(E,F)$ are free parameters of the solution. As we will see below, this case corresponds to the supersymmetric branch of solutions. The remaining solutions can be seen as limiting cases of the following possibility:

\item $\Delta_F \neq 0$

Under this assumption we can solve for $E_F$ in \eqref{eq: E_F Delta_F}:
     \begin{equation}
            E_F^2=\frac{C_F^2}{144\Delta_F} \left[36 A_F^2 - C_F^2 - 5 \right] 
            \label{eq: E_F delta neq0}
        \end{equation}
Then we see that \eqref{psaxioncpxAA} and \eqref{psaxionkAA} amount to solve the following relation:
      \begin{align}
       &\frac{8A_F^{2}C_F^{4}}{3}+4A_FC_F^{4}-\frac{7C_F^{4}}{6}+64A_F^{3}C_F^{3}+48A_F^{2}\,C_F^{3}-\frac{16A_FC_F^{3}}{3}-4C_F^{3} +576A_F^{4}C_F^{2}
       \nonumber\\ \nonumber 
           & +144A_F^{3}C_F^{2}-\frac{296A_F^{2}C_F^{2}}{3}-4A_FC_F^{2}+\frac{7C_F^{2}}{3}+2304A_F^{5}C_F -592A_F^{3}C_F+24A_F^{2}C_F\\
          & +\frac{100A_FC_F}{3}-2C_F+3456A_F^{6}-1176A_F^{4}+124A_F^{2}-\frac{25}{6} = 0\, ,
         \label{eq: megaeqF}
        \end{align}
which selects a one-dimensional family of solutions in the $(A_F,C_F)$-plane. We only consider those such that  \eqref{eq: E_F delta neq0} is non-negative, see figure \ref{fig: generalsol}.
One can check that all values in \eqref{delta=0sols} are also solutions of \eqref{eq: megaeqF}. Even if for them $\Delta_F =0$, we have that 
\be
D_F^2 = \left(1+ \frac{C_F^2(4A_F^2+1)}{\Delta_F}\right) \left[ 36A_F^2 -C_F^2-5\right]\, ,
 \label{eq: D_F delta neq0}
\ee
as well as \eqref{eq: E_F delta neq0}, attain regular limiting values that solve the equations of motion. Because \eqref{eq: megaeqF} constrains one parameter in terms of the other, we have two free parameters, say $(C,F)$, unfixed by the equations \eqref{paxionsAA} and \eqref{psaxionsAA}. 

\end{itemize}

\subsection{Summary}\label{sec:summary}

Let us summarise our results so far. Given the on-shell F-terms \eqref{solsfmax}, we find two branches of vacua, summarised in table \ref{vacuresul}. Naively, each branch seems to contain two continuous parameters. However, after choosing a specific set of flux quanta, two extra constraints will be imposed on these solutions, due to the fact that $D=m$ and eq.\eqref{extrainv}.
 Then, as we scan over different choices of flux quanta, we will obtain a discretum of values for the parameters of the Ansatz, within the above continuous solutions. In other words, the two branches become a discrete set of points once that flux quantisation is imposed.

\begin{table}[H]
\begin{center}
\scalebox{1}{%
    \begin{tabular}{| c ||     c | c | c | c |}
    \hline
  Branch & $A_F$  & $B_F$  & $C_F$  & $D_F$  \\
  \hline \hline
  \textbf{SUSY}  & $-\frac{3}{8}$ &  $-\frac{3}{2}E_F$  & $\frac{1}{4}$  & $15E_F$     \\ \hline
  \textbf{non-SUSY}   & eq.\eqref{eq: megaeqF}  & $4A_FE_F$  & eq.\eqref{eq: megaeqF}  & $ \sqrt{\frac{\Delta_F}{C_F^2}  + (4A_F^2+1)} \, 12E_F $       \\
      \hline
    \end{tabular}}      
\end{center}
\caption{Branches of solutions in terms of the quotients $A_F = A/F$, etc. of the parameters of the Ansatz \eqref{Ansatz}. In the SUSY branch $E_F$ is not constrained by the equations of motion, while in the non-SUSY extrema it is given by \eqref{eq: E_F delta neq0}. Moreover $\Delta_F$ is given by \eqref{eq: DeltaF def}, being always zero in the SUSY branch.  \label{vacuresul}}
\end{table}

As we show below, the branch where  $A_F = -3/8$, $C_F =1/4$ and $E_F$ is not constrained by the vacuum equations corresponds to supersymmetric vacua, while the other branch contains non-supersymmetric ones. Remarkably, both branches intersect at one point. The non-supersymmetric branch splits into three when imposing the physical condition $E_F^2 \geq 0$, as can be appreciated from figure \ref{fig: generalsol}. Each point of these blue curves contains two solutions, corresponding to the two values $E_F = \pm \frac{C_F}{12} \sqrt{\Delta_F^{-1}(36A_F^2 - C_F^2 - 5)}$.

\bigskip

\noindent
\textbf{F-terms}
\medskip

\noindent
One can recast the F-terms for each of these extrema as
\bes
\begin{align}
    G_{a}&=\left[\left(-\frac{1}{2}B_F-2E_F+\frac{1}{12}D_F\right)+i\left(-\frac{1}{12}C_F-\frac{1}{2}A_F-\frac{1}{6}\right)\right]F \, \mathcal{K}^2\partial_a K\, ,\\
   G_{\mu}&=\left[\left(-\frac{3}{2}B_F-\frac{1}{12}D_F-E_F\right)+i\left(-\frac{1}{4}-\frac{1}{2}A_F+\frac{1}{4}C_F\right)\right]F\, \mathcal{K}^2\partial_\mu K\, ,
\end{align}
\ees
and one can see that requiring that they vanish is equivalent to impose \eqref{SUSYsol} and \eqref{SUSYbranch}.  Therefore, the branch \eqref{SUSYsol} corresponds to supersymmetric vacua, while general solutions to \eqref{eq: megaeqF} represent non-supersymmetric extrema of the potential.

\bigskip
\noindent
\textbf{Vacuum energy and KK scale}
\medskip

\noindent
Using \eqref{eq: cosmo const relation} and imposing the extremisation of the potential, one can see that the vacuum energy has the following expression in the above branches of solutions:
\begin{equation}
   4\pi \kappa_4^4  V|_{\rm vac}= - \frac{4}{3}e^K\mathcal{K}^2F^2 \left(2A_F^2+64A_F^2E_F^2+\frac{1}{18}C_F^2+\frac{5}{18}\right)\, .
\end{equation}
In the supersymmetric branch this expression further simplifies to
\begin{equation}
    4\pi \kappa_4^4  V|_{\rm vac}^{\rm SUSY}=-  e^K\mathcal{K}^2F^2\left(12E_F^2+\frac{3}{4}\right)\, .
\end{equation}
So essentially we recover that the AdS$_4$ scale in Planck units is of order
\be
\frac{\Lambda_{\rm AdS}^2}{M_{\rm P}^2}  \sim e^{4D} V_{X_6}   F^2  \sim \frac{t^3}{u^4}  F^2 \chi\, ,
\ee
where in the last step we have defined  $\chi \equiv 2A_F^2+64A_F^2E_F^2+\frac{1}{18}C_F^2+\frac{5}{18}$. This is to be compared with the KK scale
\be
\frac{M_{\rm KK}^2}{M_{\rm P}^2} \sim   e^{2D} V_{X_6}^{-1/3} \sim  t^{-1} u^{-2} \, ,
\ee
obtaining the quotient
\be
\frac{\Lambda_{\rm AdS}^2}{M_{\rm KK}^2} \sim  e^{2D} V_{X_6}^{4/3} F^2 \sim \frac{t^{4}}{u^2} F^2 \chi \, .
\label{quot}
\ee
Scale separation will occur when this quotient is small, which seems hard to achieve parametrically, unlike in \cite{DeWolfe:2005uu,Marchesano:2019hfb}. Indeed, unless some fine tuning occurs, at large $t$, $u$ one expects that $e^K|W|^2 \sim e^K|W_{\rm RR}|^2 + e^K |W_{\rm NS}|^2$, which in supersymmetric vacua dominates the vacuum energy. If both terms are comparable, then in type IIA setups with bounded geometric fluxes and Romans mass  $u \sim t^2$, and there is no separation due to the naive modulus dependence in \eqref{quot}. If one term dominates over the other the consequences are even worse, at least for supersymmetric vacua.\footnote{With specific relations between flux quanta parametric scale separation at the 4d level is possible \cite{Font:2019uva}. Remarkably, it was there found that this naive 4d scale separation did not occur at the 10d level. \label{f:ssep}} Because $\chi$ is at least an order one number, the most promising possibility for achieving scale separation is that $F$ scales down with $t$. While this scaling is compatible with \eqref{hath}, we have not been able to find examples where this possibility is realised.\footnote{In particular, in the  SUSY branch of toroidal compactifications we have not found any flux configuration with naive scale separation beyond the case of \cite{Font:2019uva} mentioned in footnote \ref{f:ssep}.} Even if $F$ does not scale with the moduli, it would seem that generically $F \lesssim \cO(0.1)$  is a necessary condition to achieve a vacuum at  minimal scale separation. This is perhaps to be expected because in the limit $F \raw 0$ we recover the analysis of \cite{Marchesano:2019hfb}, where parametric scale separation occurs, at least from the present 4d perspective. 

In fact, the case $F=0$ also displays vacua at parametric large volume and small string coupling. While in our setup we have not been able to find families of vacua with such behaviour, one can see that small values of $F$ also favour vacua in the large volume-weak coupling regime, where the K\"ahler potential used in our analysis can be trusted. Indeed, notice that the LHS of \eqref{hath} corresponds to the contribution to the fluxes to the tadpoles and so it is a bounded integer number. As such, large values of the K\"ahler moduli will be linked to small values of $(4D_FE_F - C_F) F$. Using the scaling $u \sim t^2$, a similar conclusion can be drawn for weak coupling.

\subsection{Relation to previous results}

In order to verify the validity of our formalism and the results we have obtained, we proceed to recover some of the existing results in the literature. As argued in the next section, from the viewpoint of SU(3)-structure manifolds our vacua correspond to nearly-K\"ahler compactifications. We will therefore focus on examples that fit within that class, and mainly on two papers whose results we will link with ours.

\bigskip
\noindent
\textbf{Comparison to Camara et al. \cite{Camara:2005dc}}
\medskip

\noindent
    This reference studies RR, NS and metric fluxes on a $T^6/(\Omega(-1)^{F_L}I_3)$ Type IIA orientifold. We are particularly interested in section 4.4, where $\CN=1$ AdS vacua in the presence of metric fluxes are analysed. One can easily use our SUSY branch (see table \ref{vacuresul}), the definitions of the flux polynomials \eqref{RRrhosgeom} and our Ansatz \eqref{Ansatz} to reproduce their relations between flux quanta and moduli fixing. We briefly discuss the most relevant ones.
    
    In \cite{Camara:2005dc} they study the particular toroidal geometry in which all three complexified Kähler moduli are identified. This choice greatly simplifies the potential and the flux polynomials.  To reproduce the superpotential in \cite[eq.(3.15)]{Camara:2005dc} we consider the case $T^a=T$,  $\forall a$, so that there is only one K\"ahler modulus and the Kähler index $a$ can be removed. The flux quanta $\{e_0, e_a, m^a,m,h_\mu,\rho_{a\mu}\}$ are such that  $e_a=3c_1$, $m^a=c_2$ and 
\begin{equation}
    \rho_{a\mu}=
    \begin{cases}
    3a&\hspace{1cm} \mu=0\, ,\\
    b_\mu &\hspace{1cm} \mu\neq 0\, , 
    \end{cases}
    \qquad a, b_\mu \in \mathbb{Z}\, .
\end{equation}
Imposing the constraint $D=m$ on the SUSY Ansatz we have
\begin{align}
    A&=-\frac{3}{8}F\, ,   &    B&=-\frac{m}{10}\, , &    C&=\frac{1}{4}F\, ,    &   D&=m=15E\, .
\end{align}

        The first step is to use the invariant combinations of fluxes and axion polynomials together with the Ansatz to fix the value of the saxions. Notice that because we only have one K\"ahler modulus, $\rho_{a\mu}$ has necessarily rank one, and so \eqref{hath} fixes $t$ as function of the fluxes and the parameter $F$:
\begin{equation}
    \left(\frac{4ED}{F}-C\right)\rho_{a\mu}t^a=mh_\mu-\rho_{a\mu}m^a\longrightarrow \begin{cases} \left(\frac{4m^2}{15F}-\frac{1}{4}F\right)3a t=mh_0-3ac_2 \textrm{\hspace{1cm}if $\mu=0$}\, ,\\
    \left(\frac{4m^2}{15F}-\frac{1}{4}F\right)b_\mu t=mh_\mu-b_\mu c_2 \textrm{\hspace{1cm}if $\mu\neq0$\, .}
    \end{cases}
\end{equation}
This relation provides a constraint for the fluxes in order for this family of solutions to be realised (cf. \cite[eq.(4.32)]{Camara:2005dc}).
The complex structure saxions are instead determined in terms of  $\rho_{a\mu}$:
\begin{equation}
    \rho_{a\mu} t^a=\frac{F}{4}\mathcal{K}\partial_\mu K\longrightarrow \begin{cases}3a t=-\frac{F\mathcal{K}}{4 u^0}\, ,\\
    b_\mu t=-\frac{F\mathcal{K}}{4 u^\mu}\, ,
    \end{cases}
    \label{eq: camara u fix}
\end{equation}
which reproduces the relation  \cite[eq.(4.31)]{Camara:2005dc}.

To obtain the remaining relations of \cite[section 4.4]{Camara:2005dc}, we take into account that $\mathcal{K}=6t^3$ and take advantage of the particularly simple dependence of our Anstaz when considered on an isotropic torus. Using that $F=4C$ we can go back to \eqref{eq: camara u fix} to eliminate the $F$ dependence of the complex structure moduli. 
\begin{equation}
     \rho_{a\{\mu=0\}} t^a=F\mathcal{K}\partial_{\mu=0} K=-\frac{6t^3F}{4u^0}=-C\frac{6t^3}{u^0}=-\frac{6t^2}{u^0}\tilde{\rho}^a \longrightarrow 3atu^0=-6t^2(c_2+vm)\, ,
     \label{eq: camara u fix 2}
\end{equation}
which, up to redefinition of the parameters, is just relation  \cite[eq.(4.34)]{Camara:2005dc}. Similarly, we have
\begin{equation}
    \rho_{\mu=0}=E\mathcal{K}\partial_{\mu=0}K\longrightarrow h_0+3av=-\frac{m}{15}\frac{6t^3}{u^0}\, .
    \label{eq: camara t fix}
\end{equation}
Replacing $u_0$ using \eqref{eq: camara u fix 2} in the above expression leads to
\begin{equation}
    t^2=\frac{5(h_0+3av)(c_2+mv)}{am}\, ,
    \label{eq: camara u-t fix}
\end{equation}
which is equivalent to \cite[eq.(4.41)]{Camara:2005dc} and provides an alternative way to fix the K\"ahler moduli $t$. 

To fix the complex structure axions $\xi^\mu$ we note that
\begin{equation}
    \rho_a=B\mathcal{K}\partial_a K =-\frac{3}{2}E\mathcal{K}\partial_a K= \frac{3u^0}{2}\rho_{\mu=0}\partial_a K \longrightarrow \rho_at^a=-\frac{9}{2}(h_0+3av)u^0\, .
\end{equation}
Expanding $\rho_a$ and replacing $t$ using \eqref{eq: camara u fix 2} we arrive at
\begin{equation}
    3c_1+6c_2v+3mv^2+3a\xi^0+\sum_\mu b_\mu \xi^\mu=\frac{9}{a}(c_2+mv)(h_0+3av)\, ,
    \label{eq: complex axion combination camara}
\end{equation}
and hence we derive an analogous relation to  \cite[eq.(4.33)]{Camara:2005dc}. We observe that it  only fixes one linear combination of complex structure saxions. This was to be expected, since by construction the geometric fluxes are of rank one. Finally, we can fix the  K\"ahler axion $b$ using the flux polynomial $\rho_0$
\begin{equation}
    \rho_0=A\mathcal{K}=-\frac{3C}{2}\mathcal{K}=-\frac{3}{2t}\tilde{\rho}^a\mathcal{K}\longrightarrow\rho_0=-9(c_2+mv)t^2\, ,
\end{equation}
which after replacing the complex axions using \eqref{eq: complex axion combination camara} and substituting $t$ using \eqref{eq: camara t fix} and \eqref{eq: camara u-t fix} leads to the same equation for the K\"ahler axion as the one shown in \cite[eq.(4.40)]{Camara:2005dc}.

\bigskip
\noindent
\textbf{Comparison to Dibitetto et al. \cite{Dibitetto:2011gm}}
\medskip

\noindent
In this reference the vacuum structure of isotropic $\mathbb{Z}_2\times\mathbb{Z}_2$ compactifications is analysed, combining algebraic geometry and supergravity techniques. We are particularly interested in the results shown in \cite[section 4]{Dibitetto:2011gm}, where they consider a setup similar to \cite[section 4.4]{Camara:2005dc}, but go beyond supersymmetric vacua.\footnote{It is worth noting that in order to solve the vacuum equations, \cite{Dibitetto:2011gm} follows a complementary approach to the standard one. Typically, one starts from the assumption that the flux quanta have been fixed and then computes the values of the axions and saxions that minimise the potential. Ref.\cite{Dibitetto:2011gm} instead fixes a point in field space, and reduces the problem to find the set of consistent flux backgrounds compatible with this point being an extremum of the scalar potential. Both descriptions should be compatible.} More concretely, in this section they study type IIA orientifold compactifications on a $\mathbb{T}^6/(\mathbb{Z}_2\times \mathbb{Z}_2)$ isotropic orbifold in the presence of metric fluxes. Hence, they have an $STU$ model with the axiodilaton $S$, the overall K\"ahler modulus $T$ and the overall complex structure modulus $U$.

They obtain sixteen critical points with one free parameter and an additional solution with two free parameters. This last case is not covered by our Ansatz, since the associated geometric fluxes do not satisfy \eqref{eq: geoma} and \eqref{eq: geomu}. Therefore it should correspond to a non-supersymmetric vacuum with F-terms different from \eqref{solsfmax}. The remaining sixteen critical points are grouped into four families and summarised in \cite[table 3]{Dibitetto:2011gm}. Taking into account their moduli fixing choices, we can relate their results for the flux quanta with the parameters of our Ansatz as follows:

\begin{itemize}
    \item When $s_2=1$, solution $1$ from  \cite[table 3]{Dibitetto:2011gm} corresponds to a particular point of the SUSY branch in our table \ref{vacuresul}, with $E_F=\pm \frac{1}{4\sqrt{15}}$ (sign given by $s_1$). 
    \item When $s_2=-1$, solution $1$ of  \cite[table 3]{Dibitetto:2011gm} corresponds to the limit solution \eqref{nonSUSYdelta=0} of the non-SUSY branch (point (b) in figure \ref{fig: generalsol}). We confirm the result of \cite{Dibitetto:2011gm} regarding stability: similarly to the SUSY case, this is a saddle point with tachyonic mass $m^2=-8/9|m^2_{BF}|$ (for a detailed analysis on stability check section \ref{persta} and Appendix \ref{ap:Hessian}).
    
    \item Solution $2$ from \cite[table 3]{Dibitetto:2011gm} corresponds to a limit point $C_F=0$ of the non-SUSY branch with $\Delta_F\neq0$ and $A_F=\pm5/12$. Such solution was not detailed in our analysis of section \ref{branchvacu} since, despite being a limit point, it still verifies \eqref{eq: E_F delta neq0}, \eqref{eq: megaeqF} and \eqref{eq: D_F delta neq0}. In \cite[table 4 ]{Dibitetto:2011gm} it is stated that this solution is perturbatively unstable, in agreement with our results below (see figure \ref{fig: excludsol}).
    
    \item Solution $3$ from \cite[table 3]{Dibitetto:2011gm} is a particular case of the non-SUSY branch, corresponding to $A_F=s_1/4$ and $C_F=s_1/2$ (with $s_1=\pm1$). This specific point falls in the stable region of figure \ref{fig: excludsol}. The analysis of section \ref{persta} reveals that the mass spectrum has two massless modes, confirming the results of \cite{Dibitetto:2011gm}.

    \item Solution $4$ of \cite[table 3]{Dibitetto:2011gm} is not covered by our ansatz since, similarly to the two-dimensional solution, our parameter $F$ is not well-defined under this combination of geometric fluxes. We then expect F-terms not of the form \eqref{solsfmax}.
\end{itemize}

Hence, the results of \cite{Dibitetto:2011gm} provide concrete examples of solutions for both the supersymmetric and non-supersymmetric branches of table \ref{vacuresul}.

\bigskip
\noindent
\textbf{Examples of de Sitter extrema}
\medskip

\noindent
In \cite{Caviezel:2008tf}, the authors study the cosmological properties of type IIA compactifications on orientifolds of manifolds with geometric fluxes. They apply the no-go result of \cite{Flauger:2008ad} to rule out de Sitter vacua in all the scenarios they consider except for the manifold $SU(2)\times SU(2)$, where they find a de Sitter extremum, albeit with tachyons. One can check that the fluxes considered in section 4.2 of \cite{Caviezel:2008tf} do not satisfy condition \eqref{condhhat}. Therefore, this example lies outside of our Ansatz and so relation \eqref{eq: no-go geom inequality} does not hold. 

More generally, geometric examples of de Sitter extrema are built from compactifications on SU(3)-structure manifolds which are not nearly-K\"ahler. As we will see in section \ref{sec:10d}, our Ansatz \eqref{Ansatz} implies that the internal manifold is nearly-K\"ahler, in the approximation of smeared sources. Therefore, our analysis does not capture the attempts to find extrema in manifolds with torsion class ${\cal W}_2 \neq 0$, see e.g. \cite{Koerber:2008rx,Caviezel:2008ik,Caviezel:2008tf,Danielsson:2009ff,Danielsson:2010bc}. Remarkably, it follows from our results that such extrema cannot have F-terms of the form \eqref{solsfmax}. 


\section{Stability and 10d description}
\label{s:stabalidity}
Given the above set of 4d AdS extrema some questions arise naturally. First of all, one should check which of these points are \textit{actual} vacua, meaning stable in the perturbative sense. In other words, we should verify that they do not contain tachyons violating the BF bound \cite{Breitenlohner:1982bm}. As it will be discussed below, for an arbitrary geometric flux matrix $f_{a\mu}$ it is not possible to perform this analysis without the explicit knowledge of the moduli space metric. Nevertheless, the problem can be easily addressed if we restrict to the case in which $f_{a\mu}$ is a rank-one matrix, which will be the case studied in section  \ref{persta}. On the other hand, one may wonder if these 4d solutions have a 10d interpretation. We will see that our Ansatz can be described as an approximate SU(3)-structure background, which we will match with known 10d solutions in the literature.

\subsection{Perturbative stability}\label{persta}

 Following the approach in \cite{Marchesano:2019hfb} we will compute the physical eigenvalues of the Hessian by decomposing the K\"ahler metrics (both for the complex structure and K\"ahler fields) into their primitive and non-primitive pieces. This decomposition together with the Ansatz \eqref{Ansatz} reduces the Hessian to a matrix whose components are just numbers and whose eigenvalues are proportional to the physical masses of the moduli. The explicit computations and details are given in Appendix \ref{ap:Hessian}, whose main results we will summarise in here. To simplify this analysis we will initially ignore the contribution of the D-term potential, that is, we will set $\hat{\rho}_\alpha{}^\mu=0$. We will briefly discuss its effect at the end of this section.

As mentioned above, we will consider the case in which $f_{a\mu} = \ell_s \rho_{a\mu}$ has rank one, since the case with a higher rank cannot be solved in general. Let us see briefly why. One can show that the Ansatz \eqref{Ansatz} implies:
\begin{align}
\label{rhoproblem}
    f_{a\mu}&=-\frac{F K}{12}\p_a K\p_\mu K+\tilde f_{a\mu},    &   &\text{with}    &    & t^a \tilde f_{a\mu}=0= u^\mu \tilde f_{a\mu}\, ,
\end{align}
and so $\tilde f_{a\mu}$ must be spanned by $ t_a^\bot\otimes u_\mu^\bot$, where the  $\left\{t_a^\bot\right\}$ form a basis of the subspace orthogonal to $ t^a$, and similarly for  $ u_\mu^\bot$. The contribution of the first term of \eqref{rhoproblem} to the Hessian can be studied in general. The contribution of the second term depends, among other things, on  how both the $ t_a^{\bot}$ and $ u_\mu^{\bot}$ are stabilised, which can only be studied if the explicit form of the internal metric is known. Therefore, in the following we will set  $\tilde f_{a\mu}=0$. Notice that, for this case, our Ansatz implies that just one linear combination of axions is stabilised, since from \eqref{Ansatz} it follows that $\rho_\mu \propto \rho_{a\mu}, \forall a$.

\vspace*{.5cm}
\noindent
\textbf{SUSY Branch}
\\
As expected, the SUSY case is perturbatively stable. The results can be summarised as: 

\begin{table}[H]
\begin{center}
\scalebox{1}{%
    \begin{tabular}{| c ||     c | c | c | c |}
    \hline
  Branch & Tachyons (at least)  & Physical eigenvalues  & Massless modes (at least)  \\
  \hline \hline
  \textbf{SUSY}  & $h^{2,1}$ &  $m^2_{tach}=\frac{8}{9} m_{BF}^2$  & $h^{2,1}$  \\
  \hline
    \end{tabular}}      
\end{center}
\caption{Massless and tachyonic modes  for the supersymmetric minimum\label{susytach}.}
\end{table}
Let us explain the content of the table and especially the meaning of ``at least''. All the details of this analysis are discussed in appendix \ref{ap:Hessian}

\begin{itemize}
    \item Since the potential only depends on a linear combination of complex structure axions and the dilaton, the other $h^{2,1}$ axions of this sector are seen as flat directions. Their saxionic partners, which pair up with them into complex fields,  are tachyonic directions with mass $\frac{8}{9}m_{BF}^2$. Both modes are always present for any value of $E_F$ so we refer to them with the ``at least" tag. This is expected form general arguments, see e.g. \cite{Conlon:2006tq}. 
    
    \item For $E_F\lesssim 0.1$ there appear new tachyons with masses above the BF bound, in principle different from $\frac{8}{9}m_{BF}^2$. The masses of these modes change continuously with $E_F$, and so they become massless before becoming tachyonic.
    \item Finally, there are also modes which have a positive mass for any $E_F$.
\end{itemize} 

\vspace*{.5cm}
\noindent
\textbf{Non-SUSY branch}
\\
This case presents a casuistry that makes it difficult to summarise in just one table. As discussed in section \ref{branchvacu}, the non-SUSY vacuum candidates are described  by  the physical solutions of eq.\eqref{eq: megaeqF}, represented in figure \ref{fig: generalsol}. On top of this curve one can represent the regions that are excluded  at the perturbative level: 
\label{ss:stability}
\begin{center}    
\begin{figure}[H]
    \centering
    \includegraphics[scale=0.8]{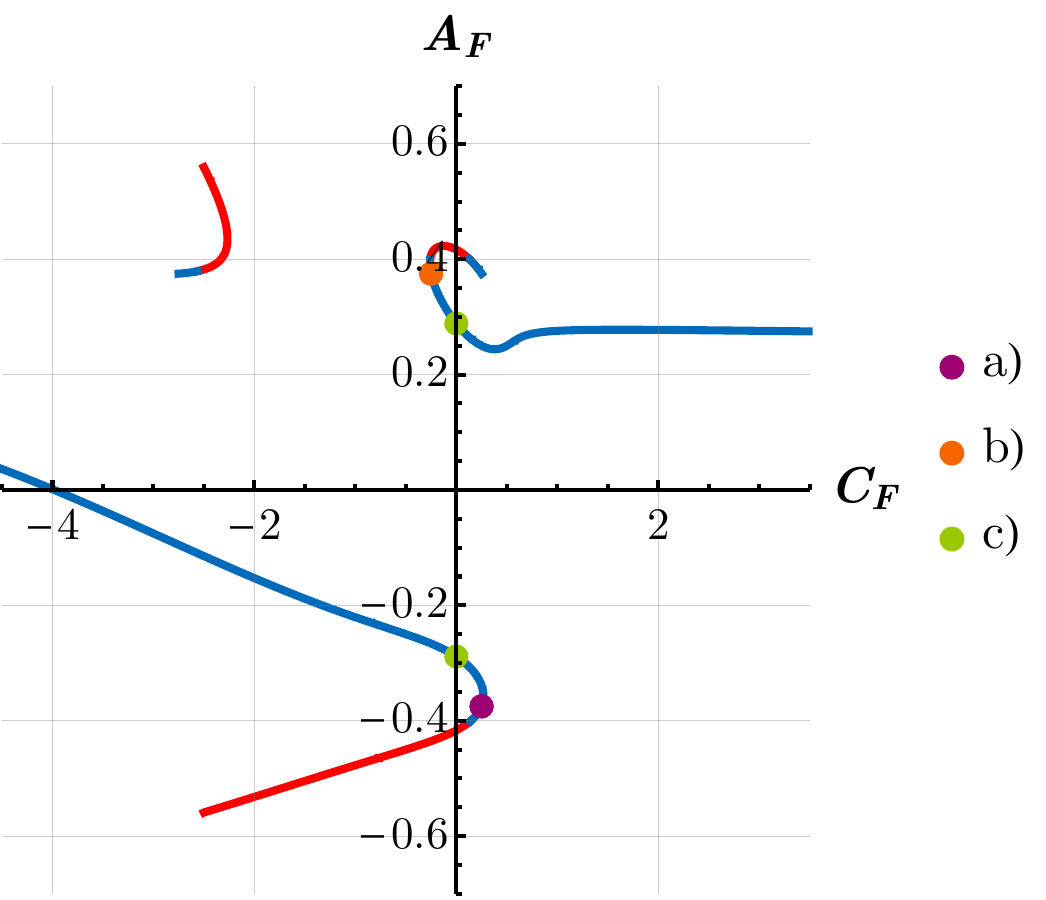}
    \caption{Set of points that verify \eqref{eq: megaeqF} with $E_F^2\geq 0$ and:  have no tachyons violating the BF bound and therefore are perturbatively  stable (blue curve);  have tachyons violating the BF bound and therefore are perturbatively unstable (red curve). The colored dots correspond to the particular solutions \eqref{delta=0sols}.}
    \label{fig: excludsol}
\end{figure}
\end{center}
Some comments are in order regarding the behaviour of the modes:
\begin{itemize}
    \item In the regions with $|A_F|\gtrsim 0.4$ there is always a tachyon whose mass violates the BF bound. This corresponds to the red pieces of the curves in figure \ref{fig: excludsol}.
    \item On the blue region of the curves, tachyons appear only in the vicinity of the red region, while away from it all the masses are positive. For instance, in the curve stretching to the right  there are no tachyons for $C_F\gtrsim 1.5$.
\end{itemize}

The explicit computation of the modes and their masses is studied in appendix \ref{ap:Hessian}.

\vspace*{.5cm}
\noindent
\textbf{D-term contribution}

As announced in the introduction, let us finish this section by commenting on the effect of the D-terms on stability. The first thing one has to notice is that, although $V_D=0$ once we impose the ansatz $\eqref{Ansatz}$, the Hessian $H_D$ associated to the D-terms is generically different from zero -see \eqref{dhessian}-. Indeed one can show that the matrix $H_D$ is a  positive semidefinite matrix. Therefore, splitting the contribution of  $V_F$ and  $V_D$ to the Hessian into $H=H_F+H_D$ and using the inequalities collected in \cite{bhatia2013matrix}, one can prove that the resulting eigenvalues of the full Hessian $H$ will always be equal or greater than the corresponding $H_F$ eigenvalues. Physically, what this means is that the D-terms push the system towards a more stable regime. In terms of the figure \eqref{fig: excludsol} and taking into account the directions affected by $H_D$ -see again \eqref{dhessian}-, one would expect that, besides having no new unstable points (red region), some of them do actually turn into stable ones (blue points) once the D-terms come into play.

\subsection{10d interpretation}
\label{sec:10d}

For those geometric vacua that fall in the large-volume regime, one may try to infer a microscopic description in terms of a 10d background AdS$_4 \times X_6$. In this section we will do so by following the general philosophy of \cite[section 5.2]{Marchesano:2019hfb}, by interpreting our 4d solution in terms of an internal manifold $X_6$ with SU(3)-structure. We hasten to stress that this does not mean that the internal metric of $X_6$ corresponds to a SU(3)-structure. As in the 10d uplift of the 4d supersymmetric vacua \cite{DeWolfe:2005uu}, recently analysed in \cite{Junghans:2020acz,Marchesano:2020qvg}, it could be that the actual 10d background displays a more general $SU(3)\times SU(3)$-structure that is approximated by an $SU(3)$-structure in some limit. This is in fact to be expected for type IIA supersymmetric backgrounds with localised sources like O6-planes, as advanced in \cite[section 5.2]{Marchesano:2019hfb}. Based on the lessons learnt from the (approximate) Calabi--Yau case \cite{Junghans:2020acz,Marchesano:2020qvg}, one should be able to describe the 4d vacua from a 10d SU(3)-structure perspective if the localised sources are smeared, so that the Bianchi identities amount to the tadpole conditions derived from \eqref{DFtadpole}, already taken into account by our analysis.

Following \cite[section 5.2]{Marchesano:2019hfb}, one may translate our Ansatz into 10d backgrounds in terms of the gauge invariant combination of fluxes
\be
G_{RR} \,=\, d_HC_{RR} + e^{-B} \wedge F_{RR} \, ,
\label{bfG}
\ee
where $d_H = d + H \wedge$. From here one reads
\begin{align}
\label{solutionsu3}
\ell_s G_6= 6 A\, d{\rm vol}_{X_6}\, , \quad \ell_s G_4 = 3 B\, J \wedge J \, , \quad \ell_s G_2 = C\, J \, , \quad \ell_s H = 6 E \,  g_s \IM (e^{-i\theta} \Omega)\, ,
\end{align}
and $ \ell_s G_0 = - D$. A vanishing D-term $D_\alpha = \frac{1}{2} \partial_\mu K \hat{f}_\alpha{}^\mu$ implies no contribution from $\hat{f}_\alpha{}^\mu$ to the torsion classes, as in the setup in \cite{Camara:2011jg}. Conversely, \eqref{eq: geoma} and \eqref{eq: geomu} imply that
\be
dJ =  \frac{3}{2}  F g_s  \ell_s \IM (e^{-i\theta} \Omega)\, , \qquad d \RE (e^{-i\theta} \Omega) = - F g_s \ell_s J \wedge J\ ,
\label{geomsu3}
\ee
which translate into the following  $SU(3)$ torsion classes
\be
{\cal W}_1 = - \ell_s g_s e^{i\theta} F \, , \qquad {\cal W}_2 = {\cal W}_3 = {\cal W}_4 = {\cal W}_5 = 0\, .
\label{torsionsu3}
\ee
Therefore, in terms of an internal SU(3)-structure manifold, our vacua correspond to nearly-K\"ahler compactifications.  

With this dictionary, it is easy to interpret our SUSY branch of solutions in terms of the general SU(3)-structure solutions for ${\cal N} = 1$ AdS$_4$ type IIA  vacua \cite{Behrndt:2004km,Lust:2004ig}. Taking for instance the choice $\theta = - \pi/2$, we can compare with the parametrisation of \cite[eq.(4.24)]{Koerber:2010bx}, and see that the relations \eqref{SUSYsol} and \eqref{SUSYbranch} fit perfectly upon identifying
\be
\ell_s |W_0|  e^{- A - i\hat{\theta}} =  3g_s\left(E + i \frac{F}{4}\right) \, ,
\label{SUSYdict}
\ee
where $|W_0|$ is the AdS$_4$ scale from the 10d frame, and $\hat{\theta}$ a phase describing the solution. 

One can in fact use this dictionary to identify some solutions in the non-supersymmetric branch with 10d solutions in the literature, like e.g. those in \cite{Lust:2008zd}. Indeed, let us in particular consider \cite[section 11.4]{Lust:2008zd}, where $\mathcal{N}=0$ AdS$_4$ compactifications are constructed by extending integrability theorems for 10d supersymmetric type II backgrounds. We first observe that the second Bianchi identity in \cite[eq.(11.29)]{Lust:2008zd} describes our first vacuum equation \eqref{paxioncpxAA}. Similarly \cite[eqs.(11.31),(11.35),(11.36)]{Lust:2008zd} are directly related to \eqref{psaxioncpxAA}, \eqref{psaxionkAA} and \eqref{paxionkAA} respectively.

Using these relations three classes of solutions are found in \cite[section 11.4]{Lust:2008zd}:
\begin{enumerate}
    \item The first solution \cite[(11.38)]{Lust:2008zd} is a particular case of the non-SUSY branch, corresponding to $A_F=\pm1/4$ and  $C_F=\pm1/2$, with $A_FC_F>0$.
    \item The second solution \cite[(11.39)]{Lust:2008zd} corresponds  the limit solution of the non-SUSY branch with $C_F=0$ and $\Delta_F\neq0$.
    \item The third solution \cite[(11.40)]{Lust:2008zd} describes a point in the SUSY branch characterised by $E_F=\pm \frac{1}{4\sqrt{15}}$. 
\end{enumerate}

To sum up, the results of \cite{Lust:2008zd} provide concrete 10d realisation of solutions for both the supersymmetric and non-supersymmetric branches of table \ref{vacuresul}.

Finally, this 10d picture allows us to understand our no-go result of section \ref{subsec: no-go's} from a different perspective. Indeed, given the torsion classes \eqref{torsionsu3} the Ricci tensor of the internal manifold $X_6$ reads \cite{BEDULLI20071125,Ali:2006gd}
\be
{\cal R}_{mn} = \frac{5}{4} g_{mn} |{\cal W}_1|^2\, ,
\ee
and so it corresponds to a manifold of positive scalar curvature, instead of the negative curvature necessary to circumvent the obstruction to de Sitter solutions \cite{Silverstein:2007ac}.


\section{Conclusions}
\label{s:conclu}

In this paper we have taken a systematic approach towards moduli stabilisation in 4d type IIA orientifold flux compactifications. The first step has been to rewrite the scalar potential, including both the F-term and D-term contributions, in a bilinear form, such that the dependence on the axions and the saxions of the compactification is factorised. This bilinear form highlights the presence of discrete gauge symmetries on the compactification, which correspond to simultaneous discrete shifts of the axions and the background fluxes. This structure has been already highlighted for the F-term piece of the potential in Calabi-Yau compactifications with $p$-form fluxes \cite{Bielleman:2015ina,Carta:2016ynn,Herraez:2018vae}, and in here we have seen how it can be extended to include general geometric and non-geometric fluxes as well. 

Besides a superpotential, these new fluxes  generate a D-term potential, which displays the same bilinear structure. The D-term potential arises from flux-induced  St\"uckelberg gaugings of the U(1)'s of the compactification by some axions that do not appear in the superpotential, and that generate conventional discrete gauge symmetries arising from $B \wedge F$ couplings. Such discrete symmetries are unrelated to the ones in the F-term potential. However, the D-term potential itself depends on the B-field axions $b^a$, because they appear in the gauge kinetic function $f_{\alpha\beta}$, and these axions do appear as well in the F-term potential, participating in its discrete symmetries. It would be interesting to understand the general structure of discrete shift symmetries that one can have in flux compactifications with both F-term and D-term potentials. In addition, it would be interesting to complete the analysis by including the presence of D6-branes with moduli and curvature corrections, along the lines of \cite{Carta:2016ynn,Herraez:2018vae,Escobar:2018tiu,Escobar:2018rna}.

As in \cite{Bielleman:2015ina,Carta:2016ynn,Herraez:2018vae}, it is the presence of discrete shift symmetries that is behind the factorisation of the scalar potential into the form \eqref{VF}, where $ Z^{\cA\cB}$ only depends on the saxionic fields, and ${\rho}_\cA$ are gauge invariant combinations of flux quanta and axions. With the explicit form of the ${\rho}_\cA$ one may construct combinations that are axion independent, and therefore invariant under the discrete shifts of the compactification. In any class of compactifications, some of the fluxes are invariant by themselves, while others need to be combined quadratically to yield a flux invariant. We have analysed the flux invariants that appear in type IIA Calabi--Yau, geometric and non-geometric flux compactifications,\footnote{The bilinear formulation may be extended to non-geometric type IIB orientifolds with O3/O7 planes along the lines of \cite{Shukla:2019wfo} which could subsequently help in performing a systematic type IIB vacua analysis.} their interest being that they determine the vev of the saxions at the vacua of the potential. Therefore, in practice, the value of these flux invariants will control whether the vacua are located or not in regions in which the effective field theory is under control. 

Another important aspect when analysing flux vacua is to guarantee their stability, at least at the perturbative level. Guided by the results of \cite{GomezReino:2006dk,GomezReino:2006wv,GomezReino:2007qi,Covi:2008ea,Covi:2008zu}, we have analysed the sGoldstino mass estimate in our setup, imposing that it must be positive as a necessary stability criterium to which de Sitter extrema are particularly sensitive. Our analysis has led us to the simple Ansatz \eqref{solsfmax} for the F-terms on-shell, which can be easily translated to relations between the $\rho_\cA$ and the value of the saxions at each extremum, cf. \eqref{proprho}.

The next step of our approach has been to find potential extrema based on this Ansatz, a systematic procedure that we have implemented for the case of geometric flux compactifications. This class of configurations is particularly interesting because they contain de Sitter extrema and are therefore simple counterexamples of the initial de Sitter conjecture \cite{Obied:2018sgi}, although so far seem to satisfy its refined version \cite{Garg:2018reu,Ooguri:2018wrx}. In this respect, we have reproduced previous de Sitter no-go results in the literature \cite{Hertzberg:2007wc,Flauger:2008ad} with our bilinear expression for the potential, but with two interesting novelties. First, when imposing that the F-terms are of the form \eqref{solsfmax} either on-shell or off-shell, we recover an inequality of the form \eqref{eq: no-go geom inequality} that forbids de Sitter extrema. We find quite amusing that this result is recovered after imposing an Ansatz inspired by de Sitter metastability. Second, our analysis includes a flux-induced D-term potential, and so the possibility of D-term uplifting, typically considered in the moduli stabilisation literature, does not seem to work in the present setting. We see our result as an interesting product of integrating several de Sitter criteria, and it would be interesting to combine it with yet other no-go results in the literature, like for instance those in \cite{Andriot:2018ept,Andriot:2019wrs,Grimm:2019ixq}. 

As is well known, type IIA  orientifold compactifications with geometric fluxes provide a non-trivial set of AdS$_4$ vacua, which we have analysed from our perspective. We have seen that, by imposing the on-shell Ansatz \eqref{solsfmax}, the equations of motion translate into four algebraic equations. By solving them, we have found two different branches of vacua, one supersymmetric and one-non-supersymmetric, and we have shown how both of them include most of the vacua found in the geometric flux compactification literature. This link with previous results can be made both with references that perform a 4d analysis and those that solve the equations of motion at the 10d level which is particularly interesting for non-supersymmetric solutions, which are scarce. Regarding 10d configurations, we have seen that our Ansatz corresponds to a nearly-K\"ahler geometry in the limit of smeared sources. This implies, in particular, that geometric flux compactifications that can be deformed to a non-trivial torsion class ${\cal W}_2$, correspond to F-terms that deviate from \eqref{solsfmax}. It would be interesting to work out the phenomenological consequences of this fact.

All these results demonstrate that analysing the bilinear form of the scalar potential provides a systematic strategy to determine the vacua of this class of compactifications, overarching previous results in the literature. Needless to say, to obtain a clear overall picture it would be important to generalise our analysis in several directions. First, it would be interesting to consider other on-shell F-term Ansatz beyond  \eqref{solsfmax} that also guarantee vacua metastability. Indeed, our analysis of the Hessian shows that, for certain geometric flux compactifications, perturbative stability occurs for a very large region of the parameter space of our F-term Ansatz, and it would be important to determine how general this result is. Second, a natural extension of our results would be to implement our  approach to compactifications with non-geometric fluxes, a task that we leave for the future. In this case it would be particularly pressing to characterise the potential corrections to the effective flux-potential, and in particular to the K\"ahler potential that we have assumed throughout our analysis. For the case of geometric fluxes these corrections should be suppressed for those vacua that sit at large volume and weak coupling. A thorough analysis of which subset of vacua lie in this regime is left for future work, although as mentioned our findings suggest that solutions with a small value for the Ansatz parameter $F$ may realise this feature.
 Remarkably, it is through the same small parameter that it seems to be possible to control the separation between the AdS$_4$ scale and the cut-off scale of the theory. This is in agreement with that in the limit $F \raw 0$ such scale separation may, a priori, be realised parametrically. 

In any event, we hope to have demonstrated that with our systematic approach one may be able to obtain an overall picture of classical type IIA flux vacua. Our strategy not only serves to find and characterise different metastable vacua, but also to easily extract the relevant physics out of them, like the F-terms, vacuum energy and light spectrum of scalars. A global picture of this sort is essential to determine what the set of string theory flux vacua is and it is not, and the lessons that one can learn from it. Hopefully, our results will provide a non-trivial step towards this final picture. 


\vspace*{1.cm}

\centerline{\bf  Acknowledgments}

\vspace*{.25cm}

We would like to thank Luis E. Ib\'a\~nez and Irene Valenzuela for useful comments and discussions.  This work is supported by the Spanish Research Agency (Agencia Estatal de Investigaci\'on) through the grant IFT Centro de Excelencia Severo Ochoa SEV-2016-0597, and by the grant PGC2018-095976-B-C21 from MCIU/AEI/FEDER, UE. D.P. is supported through the JAE Intro grant JAEINT\_19\_00948. J.Q. is supported through the FPU grant No. FPU17/04293.


\appendix


\section{Fluxes and axion polynomials}
\label{ap:conv}

In type IIA orientifold compactifications, geometric and non-geometric fluxes are defined in terms of their action on the basis of $p$-forms of table \ref{tab_1}, that correspond to the harmonic representatives of $p$-form cohomology classes of a would-be Calabi--Yau manifold $X_6$. In this framework, and following the conventions in  \cite{Ihl:2007ah}, the action of the different NS fluxes on each $p$-form is determined as 

\bea
\label{eq:fluxActions0}
& & \hskip-1.3cm H \wedge {\bf 1} = h_K \beta^K - h^\Lambda \a_\Lambda\, , \qquad \qquad \  \  H \wedge (\alpha_K + \beta^\Lambda) = -  (h_K + h^\Lambda)  \Phi_6 \nonumber\, ,\\
& &  \hskip-1.3cm f \triangleleft \om_a = f_{a K}\, \beta^K - f_a{}^\Lambda \alpha_\Lambda\, , \qquad \qquad \, f \triangleleft \varpi_\alpha = \hat{f}_{\alpha}{}^K\, \alpha_K - \hat{f}_{\alpha \, \Lambda} \beta^\Lambda\, , \nonumber\\
& &  \hskip-1.4cm f \triangleleft \alpha_K = f_{a K} \, \tilde\om^a\, , \qquad \qquad \qquad \qquad \, f \triangleleft \beta^K = -\, \hat f_{\alpha}{}^K \, \tilde\varpi^\alpha \,,\nonumber\\
& &  \hskip-1.4cm f \triangleleft \beta^\Lambda = - f_{a}{}^\Lambda  \, \tilde\om^a , \qquad \qquad \qquad \qquad   f \triangleleft \alpha_\Lambda =  \hat f_{\alpha\, \Lambda} \, \tilde\varpi^\alpha \,,\nonumber\\
& & \\
& & \hskip-1.3cm Q \triangleright \tilde\om^a = Q^{a}{}_{K}\, \beta^K- Q^{a\, \Lambda} \alpha_\Lambda\, , \qquad \quad \ \ Q \triangleright  \tilde\varpi^\alpha =  \hat{Q}^{\alpha K}\, \alpha_K- \hat{Q}^\alpha_\Lambda \beta^\Lambda \, , \nonumber\\
& & \hskip-1.3cm Q \triangleright \alpha_K = -\, Q^{a}{}_{K} \, \om_a\, , \qquad \qquad \qquad \quad  Q \triangleright \beta^K =  \hat{Q}^{\alpha\, K} \, \varpi_\alpha\,, \nonumber\\
& & \hskip-1.3cm Q \triangleright \beta^\Lambda = Q^{a\, \Lambda}  \, \om_a\, , \qquad \qquad \qquad \quad  \quad \,  Q \triangleright \alpha_\Lambda = - \hat{Q}^{\alpha}{}_\Lambda \,\varpi_\alpha\,, \nonumber\\
& & \hskip-1.3cm R \bullet {\Phi_6} = R_K\, \beta^K - R^\Lambda \alpha_\Lambda\, , \qquad \qquad \ \ R \bullet (\alpha_K + \beta^\Lambda) = ( R_K + R^\Lambda)  {\bf 1} \,, \nonumber
\eea
and we also have that $H\wedge \beta^K = H \wedge \alpha_\Lambda = R \bullet \beta^K =  R \bullet \alpha_\Lambda = 0$. The NS flux quanta are $h_K, h^\Lambda, f_{a\, K}, f_a{}^\Lambda, \hat f_{\a}{}^K, \hat f_{\a\, \Lambda}, Q^a{}_K,  Q^{a\, \Lambda}, \hat Q^{\a\, K}, \hat Q^\a{}_\Lambda, R_K, R^\Lambda \in \bZ$. This specifies the action of the twisted differential operator \eqref{eq:twistedD} on each $p$-form, and in particular the superpotential \eqref{eq:WgenNS} and the RR potential transformation \eqref{eq:C3change} leading to the D-term potential.

\subsubsection*{Axionic flux orbits and the $P$-matrices}

From the superpotential it is easy to read the gauge-invariant flux-axion polynomials \eqref{RRrhos} and \eqref{NSrhos}. Then, as in the Calabi--Yau case \cite{Herraez:2018vae}, one can check that all the remaining entries of $\rho_{\cal A}$ can be generated by taking derivatives of the {\it master polynomial} $\rho_0$. Indeed, in our more general case one finds that
\bea
\label{eq:drho0}
& & \frac{\partial\rho_0}{\partial b^a} = \rho_a\, , \quad \frac{\partial\rho_0}{{\partial b^a}{\partial b^b}} = {\cal K}_{abc} \, \tilde\rho^c\, , \quad \frac{\partial\rho_0}{{\partial b^a}{\partial b^b}{\partial b^c}} = {\cal K}_{abc} \, \tilde\rho\,, \quad \frac{\partial\rho_0}{\partial \xi^K} = \rho_K\,  , \\
& & \frac{\partial\rho_0}{{\partial b^a}{\partial \xi^K}} = \rho_{aK}\, , \quad \frac{\partial\rho_0}{{\partial b^a}{\partial b^b}{\partial \xi^K}} = {\cal K}_{abc} \, \tilde\rho^c{}_K\, , \quad \frac{\partial\rho_0}{{\partial b^a}{\partial b^b}{\partial b^c}{\partial \xi^K}} = {\cal K}_{abc} \, \tilde\rho_K \, ,\nonumber
\eea
while all the other derivatives vanish. Just like in \cite{Herraez:2018vae}, one can understand these relations from the fact that the matrix ${\cal R}$ in relating quantised and gauge invariant fluxes can be written as
\be
{\cal R}\equiv e^{b^a P_a + \xi^K P_K} \,,
\ee
with $P_a$ and $P_K$ nilpotent matrices.  Indeed, given \eqref{eq:invRmat} one can check that
\bea
\label{eq:P-matrices}
& & \hskip-0cm P_a = \begin{bmatrix}
0 & \vec\delta_a^t & 0 & 0 & 0 & 0 & 0 & 0 \\
0 & 0 & {\cal K}_{abc} & 0 & 0 & 0 & 0 & 0 \\
0 & 0 & 0 & \vec\delta_a & 0 & 0 & 0 & 0 \\
0 & 0 & 0 & 0 & 0 & 0 & 0 & 0 \\
0 & 0 & 0 & 0 & 0 & \vec\delta_a^t  \, \delta_K^L & 0 & 0 \\
0 & 0 & 0 & 0 & 0 & 0 & {\cal K}_{abc} \, \delta_K^L & 0 \\
0 & 0 & 0 & 0 & 0 & 0 & 0 & \vec\delta_a \, \delta_K^L \\
0 & 0 & 0 & 0 & 0 & 0 & 0 & 0 \\
\end{bmatrix}\, ,
\eea
and
\bea
& & P_K = \begin{bmatrix}
0 & 0 & 0 & 0 & \vec\delta_K^t & 0 & 0 & 0 \\
0 & 0 & 0 & 0 & 0 &  \vec\delta_a^t  \, \delta_K^L & 0 & 0 \\
0 & 0 & 0 & 0 & 0 & 0 &  \vec\delta_a  \, \delta_K^L & 0 \\
0 & 0 & 0 & 0 & 0 & 0 & 0 &  \vec\delta_K^t \\
0 & 0 & 0 & 0 & 0 & 0 & 0 & 0 \\
0 & 0 & 0 & 0 & 0 & 0 & 0 & 0 \\
0 & 0 & 0 & 0 & 0 & 0 & 0 &  0 \\
0 & 0 & 0 & 0 & 0 & 0 & 0 & 0 \\
\end{bmatrix}\, .
\eea

\subsubsection*{Constraints from Bianchi identities}

On compactifications with geometric and non-geometric fluxes, one important set of consistency constraints are the flux Bianchi identities. In our setup, these can be obtained by imposing that the twisted differential ${\cal D}$ in \eqref{eq:twistedD} satisfies the idempotency constraint ${\cal D}^2 = 0$ when applied on the $p$-form basis of table \ref{tab_1}  \cite{Grana:2006hr,Ihl:2007ah,Robbins:2007yv}. For simplicity let us group the 3-form/complex structure indices like in the main text, so that $(f_{aK}, f_a{}^\Lambda)$ are grouped into $f_{a\mu}$, and so on. Then, applying the definitions \eqref{eq:fluxActions0}, one obtains\footnote{Compared to \cite{Robbins:2007yv}, in our setup the flux components $h^\mu$, $R^\mu$, $f_a{}^\mu, Q^{a\mu}, \hat{f}_{\alpha \mu}$ and $\hat{Q}^\alpha{}_\mu$ are projected out. }
\bea
\label{eq:bianchids2}
& & h_\mu\, \hat{f}_\alpha{}^\mu = 0\, , \quad h_\mu\, \hat{Q}^{\alpha \mu} = 0\, , \quad f_{a\mu}\, \hat{f}_\alpha{}^\mu = 0\, , \quad f_{a\mu}\, \hat{Q}^{\alpha \mu} =0\, , \nonumber\\
& & R_\mu\, \hat{Q}^{\alpha \mu} = 0\, , \quad R_\mu \, \hat{f}_\alpha{}^\mu = 0\, , \quad Q^a{}_\mu \, \hat{Q}^{\alpha \mu} = 0\, , \quad \hat{f}_\alpha{}^{\mu} \,Q^a{}_\mu=0\, ,\\
& & \hat{f}_\alpha{}^{[\mu}\, \hat{Q}^{\alpha \nu]} = 0\, , \quad h_{[\mu} \, R_{\nu]} - f_{a[\mu}\, Q^a{}_{\nu]} = 0\,. \nonumber
\eea

\section{Curvature and sGoldstino masses} \label{ap:curvature} In this appendix we will show  that the directions \eqref{partialmax} minimise respectively $R_{a\bar c d\bar d}g^ag^bg^cg^d$ and $R_{\mu\hat\nu\rho\hat\sigma}g^{\mu}g^{\hat\rho}g^{\nu}g^{\hat\sigma}$. To do so we will follow closely \cite{Covi:2008ea,Covi:2008zu}.

\subsubsection*{Curvature}
\label{ap:curvature1}
Before talking about the extrema conditions, there are some relations that must be introduced. Consider a K\"ahler  potential depending on some set of complex chiral fields $\phi^A$ obeying a no-scale type condition:
\begin{align}
\label{noscale}
    K^A K_A=p\, ,
\end{align}
where $K_A=\nabla_A K$, $K^A=G^{A\bar B} K_{\bar B}$ and $G_{A\bar B}=\partial_A\partial_{\bar B} K$.
Taking the derivative with respect to $\nabla_B$ in \eqref{noscale} one obtains:
\begin{align}
K_B+K^A\nabla_B K_A=0\, ,
\end{align}
and deriving now with respect to $\nabla_{ C}$ we find:
\begin{align}
2\nabla_CK_B+K^A\nabla_C\nabla_B K_A=0\, . \label{hola}
\end{align}
Equation \eqref{hola} can be contracted with $K^CK^{\bar D}$ and $K^{\bar D}$  to obtain respectively
\begin{align}
R_{C\bar{D} M\bar{N}}K^CK^MK^{\bar{N}}K^{\bar D}&=\textcolor{black}{2p}\, ,&		R_{C\bar{D} M\bar{N}}K^MK^{\bar{N}}K^{\bar D}&=\textcolor{black}{2}K_C \label{curva}\, .
\end{align}
We will need these two last relations to study the extrema of $R_{A\bar B C\bar D}g^Ag^{\bar B}g^Cg^{\bar D}$

\subsubsection*{sGoldstino masses}
\label{ap:curvature2}
As discussed in section \ref{ss:fterms},
the relevant parameter to compute the sGoldstino masses is
\be
\hat{\sig} = \frac{2}{3}-R_{A\bar B C \bar D} f^{A} f^{\bar B} f^{C} f^{\bar D}\, ,
\label{app:sigma}
\ee
which we are interested in maximise. In this sense, it was shown in \cite{Covi:2008zu} that the extrema of \eqref{app:sigma} are  given  by the $f_{0A}$ satisfying the implicit relation: 
\begin{align}
    f_{0A}=\frac{R_{A\bar B C\bar D}f^{\bar B }_0f^{C}_0f^{\bar D }_0}{R_{A\bar B C\bar D}f_0^{A }f^{\bar B }_0f^{C}_0f^{\bar D }_0}\, .
    \label{app:ex}
\end{align}
Using the results above it is now straightforward to see that $f_{0A}=e^{i\alpha}\frac{K_A}{\sqrt{p}}$, $\alpha\in \mathds{R}$ are solutions of \eqref{app:ex} and therefore extrema of \eqref{app:sigma}.

\subsubsection*{Type IIA on a CY$_3$} \label{ap:curvature3}
The moduli space metric of IIA on a CY$_3$ orientifold is described  from the K\"ahler  potential:
\begin{align}
    K=K_K+K_Q\, ,
\end{align}
where the subindex $K$ refers to the K\"ahler sector whereas we use $Q$ for the complex sector. All the relations discussed above can be applied independently to $K_K$ with $p=3$ and to $K_Q$ with $p=4$. In particular, this shows that \eqref{partialmax} extremise respectively $R_{a\bar c d\bar d}g^ag^bg^cg^d$ and $R_{\mu\hat\nu\rho\hat\sigma}g^{\mu}g^{\hat\rho}g^{\nu}g^{\hat\sigma}$. Regarding the character of the points one can show that they are minima by doing small perturbations around these directions.

If one just considered the  K\"ahler sector or the complex sector (meaning taking $K_Q=0$ in the first case and $K_T=0$ in the second case) this would be the end of the story. Nevertheless, since in general we want to have both contributions, there appear some subtleties one has to take into account. The point is that now  $R_{A\bar B C\bar D}g^Ag^Bg^Cg^D$ does not have just ``one" contribution but two independent contributions:
\be
R_{A\bar B C\bar D}g^Ag^Bg^Cg^D=R_{a\bar c d\bar d}g^ag^bg^cg^d+R_{\mu\hat\nu\rho\hat\sigma}g^{\mu}g^{\hat\rho}g^{\nu}g^{\hat\sigma}\, ,
\label{app:sigma2}
\ee
and the novelty is that it new extremum appears :
\begin{align}
    f_0^A=\frac{1}{\sqrt{7}}\left\{K_a,e^{i\alpha}K_\mu\right\}
    \label{app:max}
\end{align}
with $\alpha\in\mathds{R}$, which is precisely the one discussed below \eqref{solsfmax}. Doing again a small perturbation around the points, it can be shown that now both $f_0^A=\left\{e^{i\alpha}\frac{K_a}{\sqrt{3}},0\right\}$ and $f_0^A=\left\{0,e^{i\alpha}\frac{K_\mu}{\sqrt{4}}\right\}$ are saddle points of \eqref{app:sigma2} whereas \eqref{app:max} is a  minimum.

\section{Analysis of the Hessian}
\label{ap:Hessian}

In this appendix we will compute the Hessian of the scalar potential and study its properties. We will first focus on the F-term potential, whose complexity will require a detailed analysis and the use of a simplified version of our Ansatz. Once the associated Hessian matrix has been found, we will evaluate the result in both the SUSY and the non-SUSY branches independently, in order to obtain information regarding their stability. Finally, we will briefly discuss the general behaviour of the D-term potential Hessian matrix.

\subsection*{F-term Potential}
\label{ap:Hessianf}
Starting from \eqref{eq:potentialgeom} and evaluating the second derivatives along the vacuum equations we obtain:
\bes
\begin{align}
    e^{-K}\frac{\partial^2 V_F}{\partial \xi^\sigma \partial \xi^\lambda}|_{\text{vac}}=&8\rho_\lambda\rho_\sigma+2g^{ab}\rho_{a\sigma}\rho_{b\lambda}\, ,\\
    e^{-K}\frac{\partial^2 V_F}{\partial\xi^\sigma \partial b^a}|_{\text{vac}}=&8\rho_\sigma \rho_a+8\rho_0\rho_{a\sigma}+2g^{bc}\mathcal{K}_{abd}\rho_{c\sigma}\tilde{\rho}^d\, ,\\
    e^{-K}\frac{\partial^2 V_F}{\partial \xi^\lambda \partial u^\sigma}|_{\text{vac}}=&0\, ,\\
    e^{-K}\frac{\partial^2 V_F}{\partial \xi^\sigma\partial t^a}|_{\text{vac}}=&2\partial_a g^{bc}\rho_{b\sigma}\rho_c\, ,\\
    e^{-K}\frac{\partial^2 V_F}{\partial b^a \partial b^b}|_{\text{vac}}=&8\rho_a\rho_b+8\rho_0\mathcal{K}_{abc}\tilde{\rho}^c+2g^{cd}\mathcal{K}_{ace}\mathcal{K}_{bdf}\tilde{\rho}^e\tilde{\rho}^f+2g^{cd}\mathcal{K}_{abc}\rho_d\tilde{\rho}+\frac{8\mathcal{K}^2}{9}g_{ab}\tilde{\rho}^2\nonumber\\
    &+2c^{\mu\nu}\rho_{a\mu}\rho_{b\nu}\, ,\\
    e^{-K}\frac{\partial^2 V_F}{\partial u^\sigma \partial b^a}|_{\text{vac}}=&2\partial_\sigma c^{\mu\nu}\rho_{a\mu}\rho_\nu\, ,\\
    e^{-K}\frac{\partial^2 V_F}{\partial b^a\partial t^b}|_{\text{vac}}=&2\partial_b g^{cd}\mathcal{K}_{ace}\rho_d\tilde{\rho}^e+\left(\frac{16\mathcal{K}}{3}\mathcal{K}_b g_{ac}+\frac{8\mathcal{K}^2}{9}\partial_b g_{ac}\right)\tilde{\rho}^c\tilde{\rho}\, ,\\
    \frac{\partial^2 V_F}{\partial u^\sigma\partial u^\lambda}|_{\text{vac}}=&V_F\partial_\sigma \partial_\lambda K-V_F\partial_\sigma K\partial_\lambda K\nonumber\\
    &+e^{K}\left[\partial_\sigma \partial_\lambda c^{\mu\nu}\rho_\mu\rho_\nu +t^at^b(\partial_\lambda\partial_\sigma c^{\mu\nu}\rho_{a\mu}\rho_{b\nu}-8\rho_{a\sigma}\rho_{b\lambda})+2g^{ab}\rho_{a\sigma}\rho_{b\lambda}\right]\, ,\\
    \frac{\partial^2 V_F}{\partial t^a\partial u^\sigma}|_{\text{vac}}=&V_F\partial_\sigma \partial_a K-V_F\partial_\sigma K \partial_a K+e^{K}\left[-4\mathcal{K}_a\tilde{\rho}^b\rho_{b\sigma}+4\mathcal{K}_a\tilde{\rho}\rho_\sigma\right.\nonumber\\
    &\left.-8\rho_{a\sigma}\rho_{b\mu}u^\mu t^b-8\rho_{b\sigma}\rho_{a\mu}u^\mu t^b+2\partial_\sigma c^{\mu\nu}\rho_{a\mu}\rho_{b\nu}t^b+2\partial_ag^{bc}\rho_{b\mu} u^\mu\rho_{c\sigma}\right]\, ,\\
     \frac{\partial^2 V_F}{\partial t^a\partial t^b}|_{\text{vac}}=&V_F\partial_a \partial_b K-V_F\partial_a K \partial_b K+e^{K}\left[\partial_a\partial_b g^{cd}\rho_c\rho_d+2\mathcal{K}_a\mathcal{K}_b\tilde{\rho}^2\right.\nonumber\\
     &\left.+\left(8\mathcal{K}_a\mathcal{K}_b g_{cd}+\frac{16\mathcal{K}}{3}\mathcal{K}_{ab}g_{cd}+\frac{8\mathcal{K}}{3}\mathcal{K}_a\partial_{b}g_{cd}+\frac{8\mathcal{K}}{3}\mathcal{K}_b\partial_{a}g_{cd}+\frac{4\mathcal{K}^2}{9}\partial_a\partial_bg_{cd}\right)\tilde{\rho}^c\tilde{\rho}^d\right.\nonumber\\
     &\left.+\frac{4\mathcal{K}}{3}\mathcal{K}_{ab}\tilde{\rho}^2-8\mathcal{K}_{ab}\tilde{\rho}^c\rho_{c\nu}u^\nu+8\mathcal{K}_{ab}\tilde{\rho}\rho_\nu u^\nu+2\tilde{c}^{\mu\nu}\rho_{a\mu}\rho_{b\nu}+\partial_a\partial_b g^{cd}\rho_{c\mu}\rho_{d\nu}u^\mu u^\nu\right]\, .
\end{align}
\ees
If we now introduce the ansatz \eqref{Ansatz} and make use of the decomposition of the metric in its primitive and non primitive parts -see  \eqref{eq: primitive metric}- we are left with:
\bes
\label{eq: partial second derivatives anst general}
\begin{align}
\label{ap1}
    e^{-K}\frac{\partial^2 V_F}{\partial \xi^\sigma \partial \xi^\lambda}|_{\text{vac}}=&(8E^2 +\frac{1}{6}F^2)\mathcal{K}^2\partial_\lambda K\partial_\sigma K +2g_{P}^{ab}\rho_{a\sigma}\rho_{b\lambda}\, ,\\
    e^{-K}\frac{\partial^2 V_F}{\partial \xi^\sigma \partial b^a}|_{\text{vac}}=&(8BE-\frac{4}{3}CF)\mathcal{K}^2\partial_aK\partial_\sigma K +(8A-\frac{4}{3}C)\mathcal{K}\rho_{a\sigma}\, ,\\
    e^{-K}\frac{\partial^2 V_F}{\partial \xi^\lambda \partial u^\sigma}|_{\text{vac}}=&0\, ,\\
    e^{-K}\frac{\partial^2 V_F}{\partial \xi^\sigma \partial t^a}|_{\text{vac}}=&-16B\mathcal{K}\rho_{a\sigma}\, ,\\
     e^{-K}\frac{\partial^2 V_F}{\partial b^a\partial b^b}|_{\text{vac}}=&2c^{\mu\nu}_{P}\rho_{a\mu}\rho_{b\nu}+(8B^2+\frac{4}{9}C^2+\frac{2}{9}D^2+\frac{2}{9}F^2)\mathcal{K}^2\partial_a K\partial_b K\nonumber\\
     &+(8AC-8BD-\frac{4}{3}C^2-\frac{4}{3}D^2)\mathcal{K}\mathcal{K}_{ab}\, ,\\
    e^{-K}\frac{\partial^2 V_F}{\partial b^a\partial u^\sigma}|_{\text{vac}}=&-16E\mathcal{K}\rho_{a\sigma}\, ,\\
     e^{-K}\frac{\partial^2 V_F}{\partial b^a \partial t^b}|_{\text{vac}}=&(-16BC+\frac{8}{3}CD)\mathcal{K}\mathcal{K}_{ab}\, ,\\
     e^{-K}\frac{\partial^2 V_F}{\partial u^\sigma\partial u^\lambda}|_{\text{vac}}=&(8E^2+\frac{F^2}{6})\mathcal{K}^2\partial_\sigma K\partial_\lambda K-\frac{G_{\mu\nu}}{G}(16E^2-\frac{1}{3}F^2-\frac{4}{3}DE+\frac{1}{3}CF)\mathcal{K}^2\nonumber\\
     &+2g^{ab}_{P}\rho_{a\sigma}\rho_{b\lambda} \label{eq: uu}\, ,\\
    e^{-K}\frac{\partial^2 V_F}{\partial u^\sigma\partial t^a}|_{\text{vac}}=&(-8E^2+\frac{1}{6}F^2)\mathcal{K}^2\partial_a K\partial_\sigma K-\frac{4}{3}F\mathcal{K}\rho_{a\sigma}\, ,\\
     e^{-K}\frac{\partial^2 V_F}{\partial t^a \partial t^b}|_{\text{vac}}=&(8B^2+\frac{4}{9}C^2+\frac{2}{9}D^2+\frac{2}{9}F^2)\mathcal{K}^2\partial_aK \partial_b K+ (-96B^2-\frac{8}{3}C^2+\frac{4}{3}F^2)\mathcal{K}\mathcal{K}_{ab}\nonumber\\
     &+2c^{\mu\nu}_P\rho_{a\mu}\rho_{b\nu}\label{apf}\, ;
\end{align}
\ees
where we have used the following relations
\begin{align}
    \partial_b g_{ac}t^c=-2g_{ab}\, ,\\
    \partial_\sigma\partial_\lambda c^{\mu\nu}\partial_\mu K \partial_\nu K=32 c_{\mu\nu}\, ,\\
    \partial_a\partial_b g^{cd}\partial_c K \partial_d K=32 g_{ab}\, ,\\
    \partial_a\partial_b g_{cd}t^ct^d=6g_{ab}\, .
\end{align}
Unfortunately, it is not possible to provide a general description of the stability using the results above. As discussed in section \ref{s:stabalidity}, for an arbitrary $\rho_{a\mu}$ one needs to know explicitly the internal metric. Only if we restrict ourselves to the case in which $\rho_{a\mu}$ has rank one are we able to derive a universal analysis. Therefore, from now on we will set
\begin{align}
\rho_{a\mu}&=-\frac{F}{12}\mathcal{K}\partial_aK_T \partial_\mu K_Q \label{eq: geom r1}\, .
\end{align}
Plugging this expression back into \eqref{eq: partial second derivatives anst general} the on-shell second derivatives of the potential are finally reduced to:
\bes
\begin{align}
    e^{-K}\frac{\partial^2 V_F}{\partial \xi^\sigma\partial \xi^\lambda}|_{\text{vac}}=&(8E^2+\frac{1}{6}F^2)\mathcal{K}^2\partial_\sigma K\partial_\lambda K\, ,\\
    e^{-K}\frac{\partial^2 V_F}{\partial \xi^\sigma\partial b^a}|_{\text{vac}}=&(8EB-\frac{2}{3}AF-\frac{2}{9}CF)\mathcal{K}^2\partial_\sigma K\partial_a K\, ,\\
    e^{-K}\frac{\partial^2 V_F}{\partial \xi^\sigma \partial u^\lambda}|_{\text{vac}}=&0\, ,\\
    e^{-K}\frac{\partial^2 V_F}{\partial \xi^\sigma\partial t^a}|_{\text{vac}}=&\frac{4}{3}BF\mathcal{K}^2\partial_a K\partial_\sigma K\, ,\\
    e^{-K}\frac{\partial^2 V_F}{\partial b^a\partial b^b}|_{\text{vac}}=&(8B^2+\frac{4}{9}C^2+\frac{2}{9}D^2+\frac{2}{9}F^2)\mathcal{K}^2\partial_a K\partial_b K\nonumber\\
    &+(8AC-8BD-\frac{4}{3}C^2-\frac{4}{3}D^2)\mathcal{K}\mathcal{K}_{ab}\, ,\\
    e^{-K}\frac{\partial^2 V_F}{\partial u^\sigma \partial b^a}|_{\text{vac}}=&\frac{4}{3}EF\mathcal{K}^2\partial_aK\partial_\sigma K\, ,\\
    e^{-K}\frac{\partial^2 V_F}{\partial b^a \partial t^b}|_{\text{vac}}=&(-16BC+\frac{8}{3}CD)\mathcal{K}\mathcal{K}_{ab}\, ,\\
    e^{-K}\frac{\partial^2 V_F}{\partial u^\sigma\partial u^\lambda}|_{\text{vac}}=&(8E^2+\frac{1}{6}F^2)\mathcal{K}^2\partial_\sigma K\partial_\lambda K-\frac{G_{\mu\nu}}{G}(16E^2-\frac{1}{3}F^2-\frac{4}{3}DE+\frac{1}{3}CF)\mathcal{K}^2\, ,\\
    e^{-K}\frac{\partial^2 V_F}{\partial u^\sigma\partial t^a}|_{\text{vac}}=&(-8E^2+\frac{5}{18}F^2)\mathcal{K}^2\partial_\sigma K\partial_a K\, ,\\
    e^{-K}\frac{\partial^2 V_F}{\partial t^a\partial t^b}|_{\text{vac}}=&(8A^2+16B^2+\frac{2}{9}C^2+32E^2-\frac{8}{9}F^2)\mathcal{K}^2\partial_aK\partial_b K\nonumber\, ,\\
    &+(-96B^2-\frac{8}{3}C^2+\frac{4}{3}F^2)\mathcal{K}\mathcal{K}_{ab}\, .
\end{align}
\ees
In order to make the computations manageable, we follow the same procedure as in \cite{Marchesano:2019hfb} and consider a basis of canonically normalised fields by performing the following change of basis:
\begin{align}
   \left(\xi^\mu, b^a\right)\rightarrow \left(\hat{\xi}, \hat{b}, \xi^{\hat{\mu}}, b^{\hat{a}}\right)&\, , &     \left( u^\mu, t^a\right)\rightarrow \left(\hat{u}, \hat{t}, u^{\hat{\mu}}, t^{\hat{a}}\right)& \, ,
\end{align}
where $\left\{ \hat{b}, \hat{t}\right\}$ $\left(\left\{ \hat{\xi},\hat{u}\right\}\right)$ are unit vectors along the subspace corresponding to $g_{ab}^{NP}|_{\text{vac}}$ $\left(c_{\mu\nu}^{NP}|_{\text{vac}}\right)$ and  $\left\{b^{\hat{a}}, t^{\hat{a}}\right\}$ $\left(\left\{ \xi^{\hat{\mu}},u^{\hat{\mu}}\right\}\right)$\footnote{Notice that $\hat a=1,\dots,h^{1,1}_--1$; $\hat \mu=1,\dots,h^{2,1}$} correspond analogously to vectors of unit norm with respect to $g_{ab}^{P}|_{\text{vac}}$ $\left(c_{\mu\nu}^{P}|_{\text{vac}}\right)$. We can then rearrange the Hessian $\hat H$ in a $8\times 8$ matrix with basis $(\hat{\xi}, \hat{b}, \xi^{\hat{\mu}}, b^{\hat{a}},\hat{u}, \hat{t}, u^{\hat{\mu}}, t^{\hat{a}})$ so that it reads
\begin{equation}
    \hat{H}_F=e^K\mathcal{K}^2F^2\begin{pmatrix}\frac{384{E_F}^{2}+8}{3} & H_{12} & 0 & 0 & 0 & \frac{32B}{\sqrt{3}} & 0 & 0\cr 
    H_{12} & H_{22} & 0 & 0 & \frac{32E_F}{\sqrt{3}} & H_{26} & 0 & 0\cr 
    0 & 0 & 0 & 0 & 0 & 0 & 0 & 0\cr 0 & 0 & 0 & H_{44} & 0 & 0 & 0 & H_{48}\cr 
    0 & \frac{32E_F}{\sqrt{3}} & 0 & 0 & H_{55} & H_{56} & 0 & 0\cr 
    \frac{32B_F}{\sqrt{3}} & H_{26} & 0 & 0 & H_{56} & H_{66} & 0 & 0\cr 
    0 & 0 & 0 & 0 & 0 & 0 & H_{77} & 0\cr 
    0 & 0 & 0 & H_{48} & 0 & 0 & 0 & H_{88}\end{pmatrix}\, ,
    \label{eq: hessian}
\end{equation}
where we have defined:
\begin{align}
    H_{22}=&\frac{8{D_F}^{2}-96B_FD_F+32{C_F}^{2}+96A_FC_F+864{B_F}^{2}+24}{9}\, ,\\
    H_{44}=&\frac{8{D_F}^{2}+48B_FD_F+8{C_F}^{2}-48A_FC_F}{9}\, ,\\
    H_{55}=&-\frac{192{E_F}^{2}-48D_FE_F+12C_F-20}{3}\, ,\\
    H_{66}=&\frac{3456{E_F}^{2}-8{C_F}^{2}+576{B_F}^{2}+864{A_F}^{2}-80}{9}\, ,\\
    H_{77}=&\frac{192{E_F}^{2}-16D_FE_F+4C_F-4}{3}\, ,\\
    H_{88}=&\frac{16{C_F}^{2}+576{B_F}^{2}-8}{9}\, ,\\
    H_{12}=&8\sqrt{3}\left(8B_FE_F-\frac{2C_F}{9}-\frac{2A_F}{3}\right)\\
    H_{26}=& \frac{32C_FD_F-192B_FC_F}{9}\, ,\\
    H_{48}=& -\frac{16C_FD_F-96B_FC_F}{9}\, ,\\
    H_{56}=& 8\sqrt{3}\left( \frac{5}{18}-8{E_F}^{2}\right)\, .
\end{align}
Note that \eqref{eq: hessian} defines a symmetric matrix whose components are determined once we chose a vacuum. In other words, given an extremum of the potential, one just needs to plug the correspondent $\left\{A_F,B_C,C_F,D_F\right\}$ into \eqref{eq: hessian} to analyse its perturbative stability. The physical masses of the moduli will be given by $1/2$  of the eigenvalues of the Hessian.

Once the  explicit form of Hessian has been introduced, we are ready to discuss the spectrum of the two branches obtained in the main text. This will be done in detail below.

\subsubsection*{SUSY light spectrum}
\label{sap:SUSYH}
We consider now the Hessian of the F-term potential associated to the supersymemtric branch of solutions. As explained in sections \ref{branchvacu} and \ref{sec:summary} this solution is characterised by
\begin{align}
\label{sussysol}
    A_F&=-3/8\, ,   &    B_F&=-3E_F/2\, ,   &    C_F&=1/4\, ,   &    D_F&=15E_F\, .
\end{align}
Then, one just has to plug \eqref{sussysol} into \eqref{eq: hessian}, diagonalize and divide by $1/2$ to obtain the corresponding mass spectrum. The result is:
\begin{equation}
\label{susymass}
   m^2= F^2e^K\mathcal{K}^2\left\{0,-\frac{1}{2}(1+16E_F^2),-\frac{1}{18}+56E_F^2\pm 
    \frac{1}{3}\sqrt{1+160E_F^2+2304E_F^4},\lambda_5,\lambda_6,\lambda_7,\lambda_8\right\}\, ,
\end{equation}
where the  $\lambda_i$ are the four roots of
\begin{align}
   0=&-160380+18662400E_F^2+62547240960E_F^4+2721784135680E_F^6+29797731532800E_F^8\nonumber\\
   &+(-19971-33191568E_F^2-4174924032E_F^4-74992988160E_F^6)18\lambda\nonumber\\
   &+(4483+1392480E_F^2+55800576E_F^4)\left(18\lambda\right)^2+(-133-13392E_F^2)\left(18\lambda\right)^3+\left(18\lambda\right)^4
    \label{eq: implicit}\, .
\end{align}
In order to discuss the stability, we must compare \eqref{susymass} to the BF bound, which for this case takes the value:
\begin{equation}
    m_{BF}^2=\frac{3}{4}V|_{\text{vac}}=-(\frac{9}{16}+9E_F^2)e^K\mathcal{K}^2 F^2\, .
\end{equation}
It is straightforward to see that the first non-zero eigenvalue can be rewritten as:
\begin{align}
  m_2^2= -\frac{1}{2}(1+16E_F^2)=\frac{8}{9}m_{BF}^2\, .
\end{align}
Regarding the other masses, although they can also be written as functions of $m_{BF}$ their expressions are not that illuminating. In this sense, one can check that the third eigenvalue is always positive, whereas  $m_4^2$ has a  negative region -respecting the the BF bound- for $|E_F|\lesssim 0.1$. Finally, the dependence of the four remaining eigenvalues with $E_F$, conveyed as implicit solutions of \eqref{eq: implicit}, has to be studied numerically. One finds that only one of them enters in a negative region -again above $m_{BF}^2$- for $|E_F|\lesssim 0.04$.

We conclude that the SUSY vacuum may have up to three tachyons, though only one is preserved for $|E_F| \gtrsim 0.1$. None of them violates the BF bound, as it is expected for this class of vacua. To finish this part of the appendix, let us also write the tachyonic directions:
\begin{itemize}
    \item $m_2^2$. Direction: $u^{\hat{\mu}}$.\footnote{For the complex axions, the direction $\xi^{\hat\mu}$ is the one with zero eigenvalue.}
    \item $m_4^2$.  Direction: linear combination of $b^{\hat{a}}$ and $t^{\hat{a}}$.
    \item $m_5^2=F^2e^K\mathcal{K}^2\lambda_5$ (lowest solution of \eqref{eq: implicit}). Direction: combination of all non primitive directions, i.e. $\hat{\xi}$, $\hat{b}$, $\hat{u}$ and $\hat{t}$.
\end{itemize} 

\subsubsection*{Non-SUSY branch}
\label{sap:nonSUSYH}

We end this section of the appendix by analysing the Hessian of the F-term potential associated with the non-SUSY solutions. As it was studied in detail in the main text, this branch has to be defined implicitly in terms of the $A_F$ and $C_F$ solving equation \eqref{eq: megaeqF} (check table \ref{vacuresul} and figure \ref{fig: generalsol} for details). In consequence, trying to explore the stable regions analytically is, in practice, impossible, and things must be computed numerically. What we have done is to extract  the physical $A_F$ and $C_F$ satisfying \eqref{eq: megaeqF}, plug them into \eqref{eq: hessian} -$B_F$, $D_F$ and $E_F$ are determined once $A_F$ and $C_F$ are chosen- and study the mass spectrum. Despite the numerical approach, results can be obtained easily.  

After performing a complete analysis, we conclude that a single mode is responsible for the stability of the solution. In other words, seven out of the eight masses respect the BF bound at every point of the Non-SUSY branch. Therefore, the behaviour of the aforementioned mode is precisely the one which determines the unstable region (red points) in figure \ref{fig: excludsol}. For the sake of completeness, let us write it explicitly:
\begin{align}
\label{tachmass}
  m^2=&-F^2e^K\mathcal{K}^2\left[9(12A_F^2-1)((2A_F+C_F)(6A_F+C_F)-1)\right]^{-1}\left[-9+7776A_F^6+5184A_F^5C_F\right.\nonumber\\
    &+4A_FC_F(2+C_F)(C_F^2-5C_F+9)+1296A_F^4(C_F^2-2)+144A_F^3C_F(C_F^2+C_F-9)\nonumber\\
    &\left.-C_F(C_F-2)(C_F^2+6C_F-1)+6A_F^2(C_F^4+8C_F^3-46C_F^2+4C_F+45)\right]\, .
\end{align}
As it happened in the SUSY case for the mode with mass $\frac{8}{9}m_{BF}^2$, the direction of the mode with mass \eqref{tachmass} is given by  $u^{\hat{\mu}}$. It is worth to point out that we are not saying that the other modes do not yield tachyons, but they are always above the $BF$ bound.  As discussed below figure \ref{fig: excludsol}, these other tachyons are localised close to the regions where $m^2$ defined in  \eqref{tachmass} violates the BF bound.

\subsection*{D-term potential}

We perform a similar analysis with the D-terms. Starting from \eqref{eq: D-potentialgeom} and evaluating the second derivatives along the vacuum equations, we obtain that the only non-vanishing second partial derivatives of the potential $V_D$ are
\begin{align}
     \frac{\partial^2 V_D}{\partial u^\mu \partial u^\nu}=&\frac{3}{\mathcal{K}}\partial_\mu c_{\nu\sigma}\partial_\lambda K \tilde{g}^{\alpha\beta}\hat{\rho}^\sigma_\alpha\hat{\rho}^\lambda_\beta+\frac{12}{\mathcal{K}}c_{\mu\sigma}c_{\nu\lambda} \tilde{g}^{\alpha\beta}\hat{\rho}^\sigma_\alpha\hat{\rho}^\lambda_\beta\, ,\\
   \frac{\partial^2 V_D}{\partial u^\mu \partial t^a}=&\frac{3}{\mathcal{K}}c_{\mu\sigma}\partial_\lambda K\partial_a \tilde{g}^{\alpha\beta}\hat{\rho}^\sigma_\alpha\hat{\rho}^\lambda_\beta\,-\frac{9\mathcal{K}_a}{\mathcal{K}^2}c_{\mu\sigma}\partial_\lambda K \tilde{g}^{\alpha\beta}\hat{\rho}^\sigma_\alpha\hat{\rho}^\lambda_\beta ,\\
    \frac{\partial^2 V_D}{\partial t^a \partial t^b}=&(\partial_\sigma K\partial_\lambda K\hat{\rho}^\sigma_\alpha\hat{\rho}^\lambda_\beta)\left(\frac{3}{8\mathcal{K}} \partial_a\partial_b g^{\alpha\beta}\,-\frac{9\mathcal{K}_a}{8\mathcal{K}^2} \partial_b g^{\alpha\beta}\,\right.\nonumber\\
    &\left.-\frac{9\mathcal{K}_b}{8\mathcal{K}^2} \partial_a g^{\alpha\beta}\,+\frac{27\mathcal{K}_a\mathcal{K}_b}{4\mathcal{K}^3}\partial_\sigma K\partial_\lambda K  g^{\alpha\beta}\,-\frac{9\mathcal{K}_{ab}}{4\mathcal{K}^2}  g^{\alpha\beta}\,\right) .
\end{align}
If we now take into consideration the ansatz \eqref{Ansatz} together with the Bianchi identity $f_{a\mu}\hat{f}_\alpha^\mu=0$, we have that, on-shell, $\partial_\mu K \hat{\rho}^\mu_\alpha=0$. Hence the saxionic sector of the D-term Hessian becomes
\begin{align}
    \p_A\p_B V_D=\left(\begin{matrix}\frac{12}{\mathcal{K}}c_{\mu\sigma}c_{\nu\lambda} \tilde{g}^{\alpha\beta}\hat{\rho}^\sigma_\alpha\hat{\rho}^\lambda_\beta,    &   0\\
    0   &   0
    \end{matrix}\right)\, ,
    \label{dhessian}
\end{align}
which is clearly positive-semidefinite for any choice of the geometric fluxes.


\bibliographystyle{JHEP2015}
\bibliography{reference}

\end{document}